\documentclass
[aps,twocolumn,pra,superscriptaddress,amsmath,showpacs,tightenlines,twoside]{revtex4}%
\usepackage{amsfonts}
\usepackage{amsmath}
\usepackage{amssymb}
\usepackage{version}
\usepackage[truetex]{graphicx}
\usepackage[truetex]{color}
\usepackage{hyperref}
\usepackage{float}%
\providecommand{\U}[1]{\protect\rule{.1in}{.1in}}
\begin{document}
\title{Coherent-feedback-induced photon blockade and optical bistability by an
optomechanical controller}
\author{Yu-Long Liu}
\affiliation{Institute of Microelectronics, Tsinghua University, Beijing 100084, China}
\author{Zhong-Peng Liu}
\affiliation{Department of Automation, Tsinghua University, Beijing 100084, China}
\author{Jing Zhang}
\affiliation{Department of Automation, Tsinghua University, Beijing 100084, China}
\affiliation{Tsinghua National Laboratory for Information Science and Technology (TNList),
Beijing 100084, China}
\author{Yu-xi Liu}
\email{yuxiliu@mail.tsinghua.edu.cn}
\affiliation{Institute of Microelectronics, Tsinghua University, Beijing 100084, China}
\affiliation{Tsinghua National Laboratory for Information Science and Technology (TNList),
Beijing 100084, China}
\date{\today}

\begin{abstract}
It is well-known that some nonlinear phenomena such as strong photon blockade
are hard to be observed in optomechanical system with current experimental
technology. Here, we present a coherent feedback control strategy in which a
linear cavity is coherently controlled by an optomechanical controller in a
feedback manner. The coherent feedback loop transfers and enhances quantum
nonlinearity from the controller to the controlled cavity, which makes it
possible to observe strong nonlinear effects in either linear cavity or
optomechanical cavity. More interestingly, we find that the strong photon
blockade under single-photon optomechanical weak coupling condition could be observed in the
quantum regime. Additionally, the coherent feedback loop leads to two-photon
and multiphoton tunnelings for the controlled linear cavity, which are also
typical quantum nonlinear phenomenon. We hope that our work can give new
perspectives in engineering nonlinear quantum phenomena.

\end{abstract}

\maketitle

\pagenumbering{arabic}

\section{INTRODUCTION}

Similar to Coulomb blockade for electrons in mesoscopic electronic
devices~\cite{1,2,3}, photon blockade~\cite{4,5} is a typical nonlinear
quantum optical effect, where the subsequent photons are prevented from
resonantly entering the cavity due to the strong nonlinear photon-photon
interaction. This phenomenon can be observed by the so-called photon
antibunching in photon correlation measurements. The nonlinear photon-photon
interaction at single-photon level is inherently nonclassical and provides a
way to control signal photon via photonic devices, which is essential to
various emerging techniques, such as single-photon transistor~\cite{6}, photon
routing~\cite{7,8,9,10,11}, generation of non-classical light~\cite{12},
quantum information processing with photonic qubits~\cite{13,14,16}, optical
communication~\cite{17}, and optical quantum computer~\cite{18}. The photon
blockade has been experimentally demonstrated in, e.g., cavity-QED systems
with the strong atom-cavity coupling~\cite{19}, a quantum dot strongly coupled
to a photonic crystal resonator~\cite{20}, and circuit-QED
systems~\cite{21,22}. It is generally recognized that the nonlinear photon
coupling strength should be far larger than the cavity decay rate when the
single-photon blockade occurs. Thus engineering nonlinear photon-photon
coupling is an important task for single-photon devices.

Recently, cavity optomechanics becomes a rapidly growing field of research, in
which nonlinear couplings between the electromagnetic and mechanical degrees
of freedom~\cite{23,24,Clerk,25,26} lead to various interesting phenomena. The
nonlinear optomechanical coupling can be applied to the detection of
gravitational waves~\cite{27,28,29}, the observation of quantum effects in the
mescoscopic and macroscopic scales~\cite{30}. The optomechanical system can
also be used to build sensitive mass, force and displacement
detectors~\cite{31}, and the hardware for realizing quantum information
processing~\cite{32}. Various achievements have been made both theoretically
and experimentally in optomechanical systems, for instance, the cooling of the
mechanical resonator to its ground state, which paves the way to study the
physical effects on the boundary between classical and quantum
mechanics~\cite{33,34,35,36,37,38,39,40,41,42,43,44,45,46,47}, the mechanical
oscillations induced by the radiation pressure~\cite{48,49},
electromagnetically induced transparency~\cite{50,51,52,53,54}, entanglement
between optical and mechanical modes~\cite{55,56}, optomechanical
transducers~\cite{57}, normal mode splitting~\cite{58}, and coherent optical
wavelength conversion~\cite{59,60,61,62,HLWang,63,Lehnert}. Especially, the
nonlinear Kerr effect can also be realized in optomechanical
systems~\cite{Gong,80,Lu}, which can be used to engineer nonlinear
photon-photon interaction. Thus, the optomechanical systems might also be very
important candidates to act as single-photon devices, e.g., a single-photon
router~\cite{AgarwalPRA85}.

It has been theoretically shown that the single-photon phenomena and photon
blockade can be realized in optomechanical
systems~\cite{64,65,83,Jieqiao1,Jieqiao2} when the single-photon
optomechanical coupling strength is much larger than the cavity decay rate.
However, in current experimental technology, the optomechanical-coupling
induced Kerr nonlinearity is not strong enough to be used for observing such
single-photon effect. Several approaches have been proposed for enhancing the
photon-photon interaction using coupled optomechanical systems~\cite{66,67}.
However, the strong single-photon optomechanical coupling is still a necessary
condition for the demonstration of photon blockade. Recent studies
showed~\cite{68} that particular nonlinear effects are hopeful to be observed
in a linear cavity, which is coupled to an optomechanical system with the weak
optomechanical coupling. However, photon blockade is still hard to be observed
in optomechanical system. Moreover, this method requires strong coupling
between the linear cavity and the optomechanical cavity via spatial proximity.
However, individual addressability of each cavity is still an arduous
challenge in current experimental technology~\cite{Arka majumdar}.

We recently study a method to induce nonlinearity into a linear cavity by
coherent feedback control using a circuit QED system as a
controller~\cite{78,Z.P.L}. Motivated by
studies~\cite{Gong,80,Lu,AgarwalPRA85,64,65,83,Jieqiao1,Jieqiao2,66,67,68} and
considering progress on quantum coherent feedback
methods~\cite{Feedback,69,70,71,72,73,74,75,76,FeedbackExp1,FeedbackExp2,FeedbackExp3,77}%
, we propose an approach to realize an optomechanically-based single-photon
device by replacing the circuit QED system in the coherent feedback
approach~\cite{Z.P.L} with a traditional optomechanical system. Such a
coherent-feedback strategy can liberate both the linear cavity and the
optomechanical system in space and eliminate the slashing requirement on
individual addressability or large coupling strength between linear cavity and
optomechanical cavity. Moreover, we find that photon blockade phenomenon can
be observed even in the weak optomechanical coupling regime for such a design.
We will also show that the damping effects of the controlled cavity and the
optomechanical controller are actually crucial for achieving photon blockade
in such coherent-feedback approach. This is different from other existing
photon blockade systems in which the cavity loss always plays a negative role.

In our approach, the output of the controlled cavity is unitarily processed by
optomechanical controller, and then the processed output field is fed into the
controlled cavity again. Such a coherent-feedback strategy preserves the
quantum coherence of the system and also reduces the feedback-induced time
delay. In contrast to direct coherent feedback~\cite{69}, we use
field-mediated coherent feedback~\cite{72,73,74} method, in which the
information flow is uniquely determined by the propagation direction of the
quantum field and thus it is easier to be realized in experiments.

The paper is organized as follows. In Sec.~II, we briefly summarize the main
results of quantum input-output and coherent feedback control theory which are
related to our study. In Sec. III, we present the mathematical descriptions of
our feedback control system, i.e., a linear cavity coupled to an
optomechanical system in the feedback configuration. We introduce both the
steady-state equations and the quantum Langevin equations with quantum and
thermal fluctuations to model the dynamics of our feedback control system. In
Sec. IV, we study the nonlinear effects of the controlled linear cavity
induced by the optomechanical system in the semi-classical regime. The
nonlinear effects of the controlled cavity, such as optical bistability, are
induced by the nonlinear dissipative coupling between the controlled system
and the intermediate quantum field. In Sec. V, we studied the statistical
properties of photons in both the optomechanical controller and the controlled
cavity with two different driving methods in the quantum regime. Our results
show that the strong photon blockade effect can be observed even in the weak
single-photon optomechanical coupling regime. Feedback-induced photon
tunneling processes, especially two- and three-photon tunnelings, are also
discussed. Conclusions and perspective discussions are given in Sec. VI.

\section{Field-MEDIATED COHERENT FEEDBACK}

\begin{figure}[h]
\includegraphics[bb=0 0 562 231, width=5cm, clip]{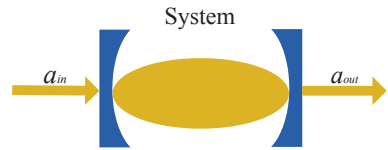} \caption{(Color
online) Schematic diagram of a single quantum system with the input field
$a_{\mathrm{in}}$ and output field $a_{\mathrm{out}}$.}%
\label{fig1}%
\end{figure}

The quantum system with an input field $a_{\mathrm{in}}$ and an output field
$a_{\mathrm{out}}$ can be schematically shown in Fig.~\ref{fig1}. The input
field $a_{\mathrm{in}}$ can be described by a continuum of harmonic
oscillators. Under the Markovian approximation, the Hamiltonian $H$ of the
whole system can be given as~\cite{81}
\begin{equation}
H=H_{\mathrm{sys}}+i[a_{\mathrm{in}}^{\dag}L-L^{\dag}a_{\mathrm{in}}].
\label{eq:1}%
\end{equation}
Here, we assume $\hbar=1$. We also use $H_{\mathrm{sys}}$ to denote the
Hamiltonian of the quantum system. $L=\sqrt{\gamma}a$ is the Lindblad operator
induced from the system and the bath field, where $a\,(a^{\dag})$ is the
annihilation (creation) operator of the system. The input field
$a_{\mathrm{in}}$ is defined as
\begin{equation}
a_{\mathrm{in}}=\frac{1}{\sqrt{2\pi}}\int_{-\infty}^{+\infty}b(\omega
)e^{-i\omega t}d\omega.
\end{equation}
where $b(\omega)\,(b^{\dag}(\omega))$ is the annihilation (creation) operator
of the bath mode with frequency $\omega$ satisfying $[b(\omega),b^{\dag
}(\omega^{^{\prime}})]=\delta(\omega-\omega^{^{\prime}})$ and $[b(\omega
),b(\omega^{^{\prime}})]=0$.

The system shown in Fig.~\ref{fig1} can also be modelled by a set of
parameters $G=(S,L,H)$~\cite{73}. Here, $S$ denotes the scattering matrix, $H$
and $L$ are given in Eq.~(\ref{eq:1}). The $(S,L,H)$ notation can be used
conveniently to study the networks of coupled open quantum systems for quantum
control analysis and design.

\begin{figure}[h]
\includegraphics[bb=0 0 585 445,  width=8.5cm, clip]{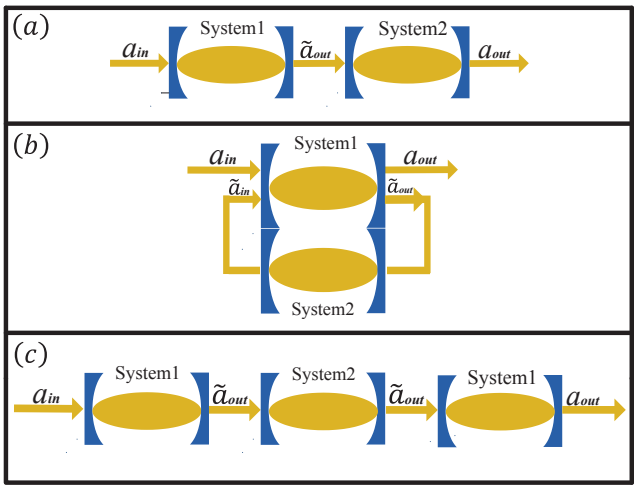} \caption{(Color
online) Schematic diagrams of the cascade quantum system. (a) Cascade system
with two cascaded-connected components, in which the output of the first
system acts as the input field of the second system. (b) Proposed coherent
feedback system, in which the output of the first system is fed into the
quantum controller coherently and then the output of the quantum controller is
coherently fed back to act as the input of the first system. (c) Equivalent
schematic of coherent feedback system, which can be seen as the controlled
system cascaded-connected with the controller system and then
cascaded-connected with itself.}%
\label{fig2}%
\end{figure}

The theory of the single quantum system with input and output fields, as shown
in Fig.~\ref{fig1}, can be generalized to a Markovian quantum cascaded system
as shown in Fig.~\ref{fig2}(a). We assume that the output field of the first
system, described by $G_{1}=(S_{\mathrm{1}},L_{\mathrm{1}},H_{\mathrm{1}})$,
acts as the input field of the second system, described by $G_{2}%
=(S_{\mathrm{2}},L_{\mathrm{2}},H_{\mathrm{2}})$. This coupled cascade system
is equivalent to the system, described by~\cite{73}%

\begin{equation}
G=(S_{\mathrm{eff}},H_{\mathrm{eff}},L_{\mathrm{eff}}),
\end{equation}
with%
\begin{align}
S_{\mathrm{eff}}  &  =S_{\mathrm{2}}S_{\mathrm{1}},\text{ }L_{\mathrm{eff}%
}=L_{\mathrm{2}}+S_{\mathrm{2}}L_{\mathrm{1}},\\
H_{\mathrm{eff}}  &  =H_{\mathrm{1}}+H_{\mathrm{2}}+\frac{1}{2i}%
(L_{\mathrm{2}}^{\dag}S_{\mathrm{2}}L_{\mathrm{1}}-L_{\mathrm{1}}^{\dag
}S_{\mathrm{2}}^{\dag}L_{\mathrm{2}}).
\end{align}

As shown in Fig.~\ref{fig2}(b), we now focus on another scenario in which the
output of the first system $G_{\mathrm{1}}=(S_{\mathrm{1}},L_{\mathrm{1}%
},H_{\mathrm{1}})$ is taken as the input of the second system with
$G_{\mathrm{2}}=(S_{\mathrm{2}},L_{\mathrm{2}},H_{\mathrm{2}})$, and
simultaneously the output of the second system is taken as the input of the
first system, by which a coherent-feedback loop is constructed. In
Fig.~\ref{fig2}, both the scattering matrices of these two components are
identity matrix $I$, that is, $S_{\mathrm{1}}=S_{\mathrm{2}}=S=I$. Such
coherent-feedback system is equivalent to a system in which $G_{\mathrm{1}%
}=(I,L_{\mathrm{1}},H_{\mathrm{1}})$ is first cascaded-connected to
$G_{\mathrm{2}}=(I,L_{\mathrm{2}},H_{\mathrm{2}})$ and then cascade-connected
to $G_{\mathrm{3}}=(I,L_{\mathrm{3}},H_{\mathrm{1}})$ as shown in
Fig.~\ref{fig2}(c). The whole feedback system, shown in both Fig.~\ref{fig2}%
(b) and Fig.~\ref{fig2}(c), can be described by%

\begin{equation}
(\tilde{S},\tilde{L},\tilde{H}),
\end{equation}
with
\begin{align}
\tilde{S}  &  =S,\text{ }\tilde{L}=L_{\mathrm{1}}+L_{\mathrm{2}}%
+L_{\mathrm{3}},\\
\tilde{H}  &  =H_{\mathrm{1}}+H_{\mathrm{2}}+\frac{1}{2i}\left[
(L_{\mathrm{2}}^{\dag}L_{\mathrm{1}}+L_{\mathrm{3}}^{\dag}L_{\mathrm{2}%
}+L_{\mathrm{3}}^{\dag}L_{\mathrm{1}})-\mathrm{H.C.}\right]  .
\end{align}

\section{MODEL AND HAMILTONIAN}

\begin{figure}[h]
\includegraphics[bb=0 0 560 520,  width=0.3\textwidth, clip]{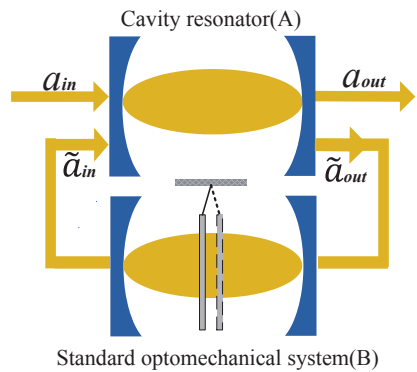}\caption{(Color
online) Schematic diagrams for a cavity (A) coherently feedback controlled by
a standard optomechanical system (B). }%
\label{fig3}%
\end{figure}

As schematically shown in Fig.~\ref{fig3}, we study a linear cavity which is
controlled by a standard optomechanical system. The output field of the
controlled cavity is coherently fed into the optomechanical controller and
then fed back into the controlled cavity again. The controlled cavity can be
taken as a transmission line resonator, a toroidal microresonator, a cavity
with two mirrors, or a defect cavity in photonic crystal. Without loss of
generality and for simplicity, we will focus on the cavity with two mirrors
which can support two input channels and two output channels.
The dissipation of the controlled cavity via the vacuum
fluctuation field $a_{\mathrm{in}}$ is described by the Lindblad operator
$L_{\mathrm{a}}=\sqrt{\kappa}a$. $a\,(a^{\dag})$ is the annihilation
(creation) operator of the controlled cavity with the decay rate $\kappa$ and
the frequency $\omega_{\mathrm{s}}$. $H_{\mathrm{a}}=\omega_{\mathrm{s}%
}a^{\dag}a$ describes the Hamiltonian of the controlled cavity. With the
$(S,L,H)$ notation, the controlled cavity can be described by
$(S,L_{\mathrm{a}},H_{\mathrm{a}})$. The output of the controlled cavity is
fed into a standard optomechanical system, which serves as a controller to
induce and manipulate the nonlinearity of the controlled cavity. The resonant
frequency of the cavity of optomechanical system is modulated by the position
of a mechanical resonator. A monochromatic coherent light field with the
frequency $\omega_{\mathrm{d}}$ and amplitude $\epsilon$ is used to drive the
cavity of the optomechanical system. The driven Hamiltonian $H_{\mathrm{c}}$
of the optomechanical system is given by%

\begin{equation}
H_{\mathrm{c}}=\omega_{\mathrm{c}}c^{\dag}c+\omega_{\mathrm{m}}b^{\dag
}b+g_{_{\mathrm{0}}}c^{\dag}c(b^{\dag}+b)+\epsilon(c^{\dag}e^{i\omega
_{\mathrm{d}}t}+ce^{-i\omega_{\mathrm{d}}t}),
\end{equation}
where $c\,(c^{\dag})$ is the annihilation (creation) operator of the
optomechanical cavity, $b\,(b^{\dag})$ is the annihilation (creation) operator
of the mechanical model with frequency $\omega_{\mathrm{m}}$. The parameter
$g_{_{\mathrm{0}}}$ is the single-photon optomechanical coupling strength. The
optomechanical controller interacts with the intermediate field via the
dissipation channel of the cavity field described by the Lindblad operator
$L_{\mathrm{c}}=\sqrt{\gamma}c$. Afterward, the output of the controller is
fed into the controlled system via the dissipation channel $L_{\mathrm{f}%
}=\sqrt{\kappa_{\mathrm{f}}}a$ to complete the whole coherent feedback loop.
With the $(S,L,H)$ notation, the optomechanical controller can be described by
$(S,\sqrt{\gamma}c,H_{\mathrm{c}})$.

The whole coherent feedback system can be described by three
cascaded-connected subsystems which have been schematically shown in
Fig.~\ref{fig2}(b). Thus, the cascade system, with the first ($G_{\mathrm{1}%
}=(I,\sqrt{\kappa}a,\omega_{\mathrm{s}}a^{\dag}a)$), the second
($G_{\mathrm{2}}=(I,\sqrt{\gamma}c,H_{\mathrm{c}})$), and, the third
($G_{\mathrm{3}}=(I,\sqrt{\kappa_{\mathrm{f}}}a,\omega_{\mathrm{s}}a^{\dag}%
a)$) subsystems, can be described by%

\[
(S^{^{\prime}},L^{^{\prime}},H^{^{\prime}}).
\]
with%

\begin{align}
S^{^{\prime}}  &  =I,L^{^{\prime}}=\sqrt{\kappa}a+\sqrt{\gamma}c+\sqrt
{\kappa_{\mathrm{f}}}a,\\
H^{^{\prime}}  &  =H_{\mathrm{a}}+H_{\mathrm{c}}+\frac{i}{2}(\sqrt
{\gamma\kappa}-\sqrt{\gamma\kappa_{\mathrm{f}}})(a^{\dag}c-c^{\dag}a).
\label{Htot}%
\end{align}
In the rotating reference frame with unitary transformation $R(t)=\exp
[i\omega_{\mathrm{d}}(c^{\dag}c+a^{\dagger}a)t]$, the Hamiltonian in
Eq.~(\ref{Htot}) becomes%

\begin{align}
\tilde{H}  &  =\Delta_{\mathrm{s}}a^{\dag}a+\Delta_{\mathrm{c}}c^{\dag
}c+\omega_{\mathrm{m}}b^{\dag}b+g_{_{\mathrm{0}}}c^{\dag}c(b^{\dag
}+b)\nonumber\\
&  +\epsilon(c^{\dag}+c)+\frac{i}{2}(\sqrt{\gamma\kappa}-\sqrt{\gamma
\kappa_{\mathrm{f}}})(a^{\dag}c-c^{\dag}a),
\end{align}
where $\Delta_{\mathrm{s}}=\omega_{\mathrm{s}}-\omega_{\mathrm{d}}$ and
$\Delta_{\mathrm{c}}=\omega_{\mathrm{c}}-\omega_{\mathrm{d}}$ are the detuning frequencies.

Using the input-output theory, the dynamics of the whole system is described
by the quantum Langevin equation(QLEs)~\cite{Z.P.L,77,78}%
\begin{align}
\dot{a}  &  =-i\Delta_{\mathrm{s}}a-\frac{1}{2}(\sqrt{\kappa}+\sqrt
{\kappa_{\mathrm{f}}})^{2}a-\sqrt{\gamma\kappa_{\mathrm{f}}}c\nonumber\\
&  -(\sqrt{\kappa}+\sqrt{\kappa_{\mathrm{f}}})a_{\mathrm{in}},\\
\dot{b}  &  =-i\omega_{\mathrm{m}}b-ig_{_{\mathrm{0}}}c^{\dag}c-\frac
{\gamma_{\mathrm{m}}}{2}b-\sqrt{\gamma_{\mathrm{m}}}b_{\mathrm{in}},\\
\dot{c}  &  =-i\Delta_{\mathrm{c}}c-ig_{_{\mathrm{0}}}c(b^{\dag}%
+b)-\sqrt{\kappa\gamma}a-\frac{1}{2}\gamma c\nonumber\\
&  -i\epsilon-\sqrt{\gamma}a_{\mathrm{in}},
\end{align}
where $\gamma_{\mathrm{m}}$ is the damping rate of the mechanical resonator.
$b_{\mathrm{in}}$ denotes the thermal noise acting on the mechanical resonator
which satisfies the Markovian correlation relation
\begin{equation}
\langle b_{\mathrm{in}}(t)b_{\mathrm{in}}^{\dag}(t^{^{\prime}})\rangle
=(n_{\mathrm{th}}+1)\delta(t-t^{^{\prime}}),
\end{equation}
The mean thermal occupation number of $b_{\mathrm{in}}$ can be calculated by
$n_{\mathrm{th}}=[\exp(\omega_{m}/k_{\mathrm{B}}T)-1]^{-1}$. We also assume
that the vacuum fluctuation $a_{\mathrm{in}}$ satisfies
\begin{equation}
\langle a_{\mathrm{in}}(t)a_{\mathrm{in}}^{\dag}(t^{^{\prime}})\rangle
=\delta(t-t^{^{\prime}}).
\end{equation}
where the frequency of the controlled cavity is assumed to be much higher than
that of the mechanical resonator, thus the temperature effect on the cavity
field has been neglected.

Using the mean field approximation, the time evolutions of the mean values of
each operator can be given as:
\begin{align}
\frac{d\left\langle a\right\rangle }{dt}  &  =-i\Delta_{\mathrm{s}%
}\left\langle a\right\rangle -\frac{1}{2}(\sqrt{\kappa}+\sqrt{\kappa
_{\mathrm{f}}})^{2}\left\langle a\right\rangle \label{meana}\\
&  -\sqrt{\gamma\kappa_{\mathrm{f}}}\left\langle c\right\rangle ,\nonumber\\
\frac{d\left\langle b\right\rangle }{dt}  &  =-i\omega_{\mathrm{m}%
}\left\langle b\right\rangle -ig_{_{\mathrm{0}}}\left\vert c\right\vert
^{2}-\frac{\gamma_{\mathrm{m}}}{2}\left\langle b\right\rangle ,\\
\frac{d\left\langle c\right\rangle }{dt}  &  =-i\Delta_{\mathrm{c}%
}\left\langle c\right\rangle -ig_{_{0}}\left\langle c\right\rangle
(\left\langle b\right\rangle ^{\ast}+\left\langle b\right\rangle
)-i\epsilon\label{meanc}\\
&  -\sqrt{\kappa\gamma}\left\langle a\right\rangle -\frac{1}{2}\gamma
\left\langle c\right\rangle .\nonumber
\end{align}

\section{COHERENT FEEDBACK INDUCED OPTICAL BISTABILITY}

It is known that the bistability can be found in the standard optomechanical
system~\cite{80}. Let us now first study how the optical nonlinear behavior in
the controlled linear cavity can be induced by a standard optomechanical
system using coherent feedback method. In other words, we study how the
optical nonlinearity in optomechanical system can be manipulated and
transferred to a linear cavity by using coherent feedback. By solving
Eqs.~(\ref{meana}-\ref{meanc}) with $\langle\dot{a}\rangle_{\mathrm{s}%
}=\langle\dot{b}\rangle_{\mathrm{s}}=\langle\dot{c}\rangle_{\mathrm{s}}=0$, we
can obtain the steady-state values
\begin{align}
A_{\mathrm{0}}  &  =\frac{-\sqrt{\gamma\kappa_{\mathrm{f}}}}{[i\Delta
_{\mathrm{s}}+\frac{1}{2}(\sqrt{\kappa}+\sqrt{\kappa_{\mathrm{f}}})^{2}%
]}C_{\mathrm{0}},\label{A0}\\
B_{\mathrm{0}}  &  =\frac{-ig_{_{\mathrm{0}}}}{i\omega_{\mathrm{m}}%
+\frac{\gamma_{\mathrm{m}}}{2}}\left\vert C_{\mathrm{0}}\right\vert
^{2},\label{B0}\\
C_{\mathrm{0}}  &  =\frac{-i\epsilon-\sqrt{\kappa\gamma}A_{\mathrm{0}}%
}{i\Delta_{\mathrm{c}}+ig_{_{\mathrm{0}}}(B_{\mathrm{0}}^{\ast}+B_{\mathrm{0}%
})+\frac{1}{2}\gamma}, \label{C0}%
\end{align}
of the cavity fields and mechanical resonator. Here, $A_{\mathrm{0}%
}=\left\langle a\right\rangle _{\mathrm{s}}$, $B_{\mathrm{0}}=\left\langle
b\right\rangle _{\mathrm{s}}$, and $C_{\mathrm{0}}=\left\langle c\right\rangle
_{\mathrm{s}}$ denote the steady state values of the average $\left\langle
a\right\rangle $, $\left\langle b\right\rangle $ and $\left\langle
c\right\rangle $.

\begin{figure}[t]
\includegraphics[bb=0 192 570 632,width=0.25\textwidth, clip]{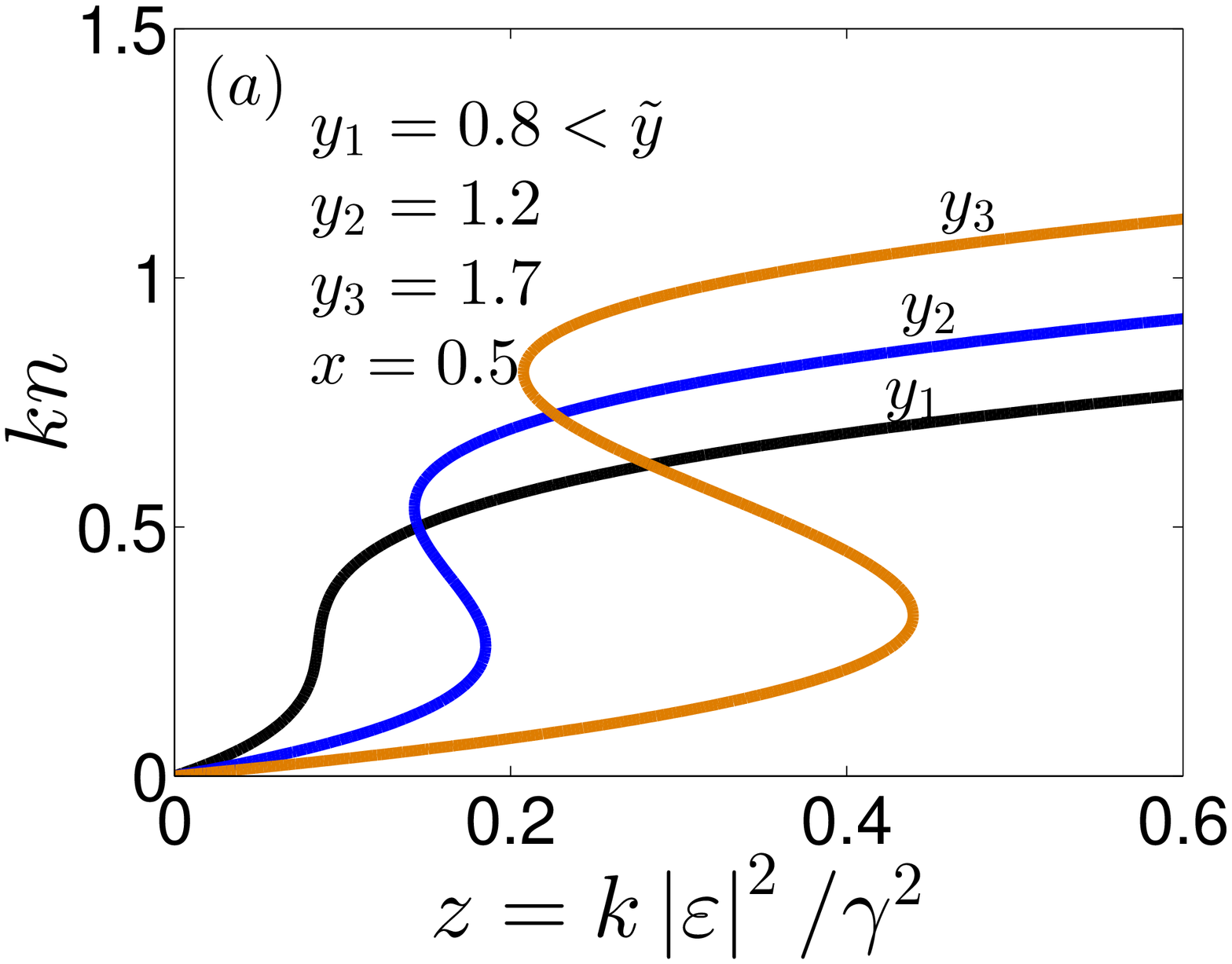}\includegraphics[bb=0 192 570 632,width=0.25\textwidth, clip]{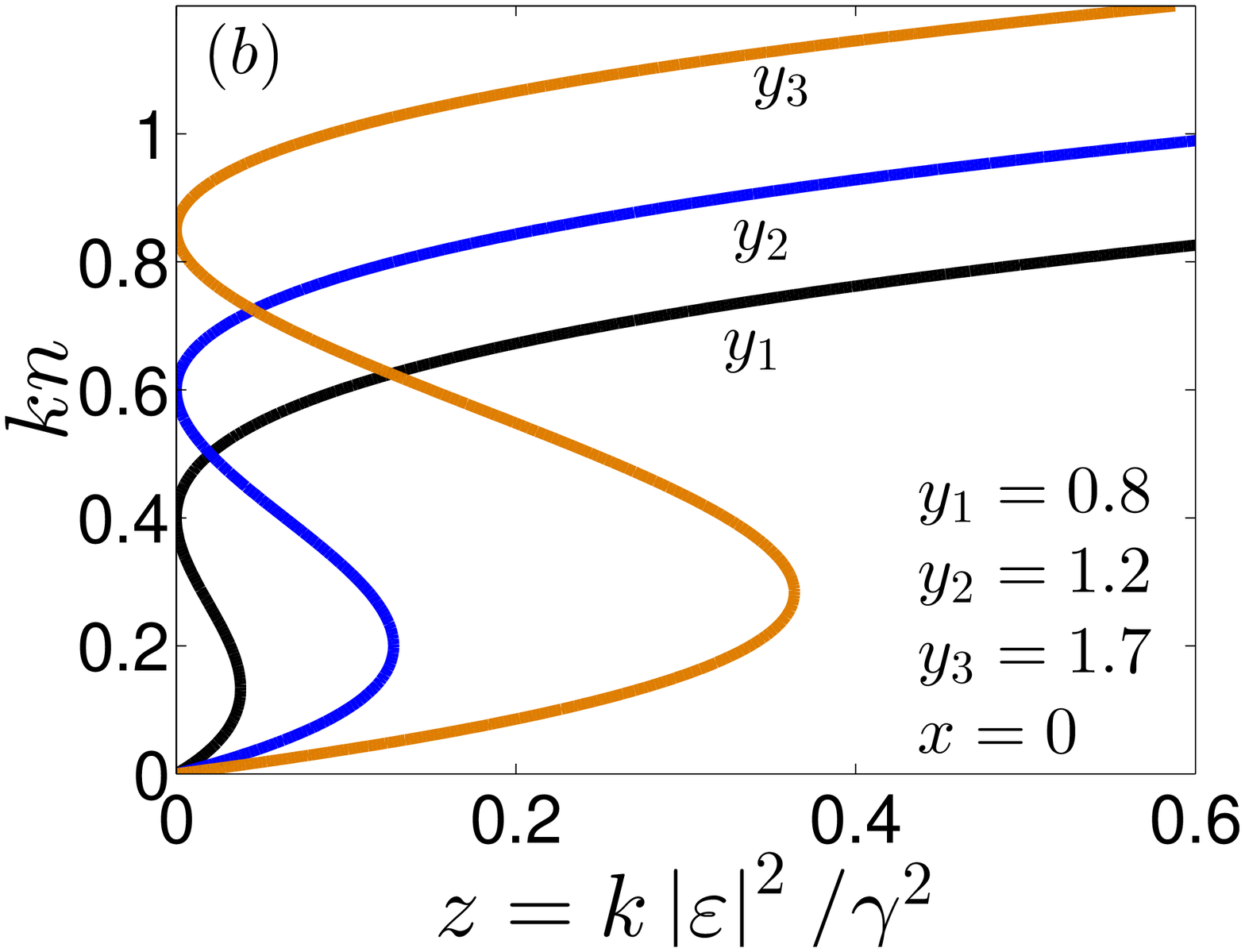}\newline%
\includegraphics[bb=0 192 570 632,width=0.25\textwidth, clip]{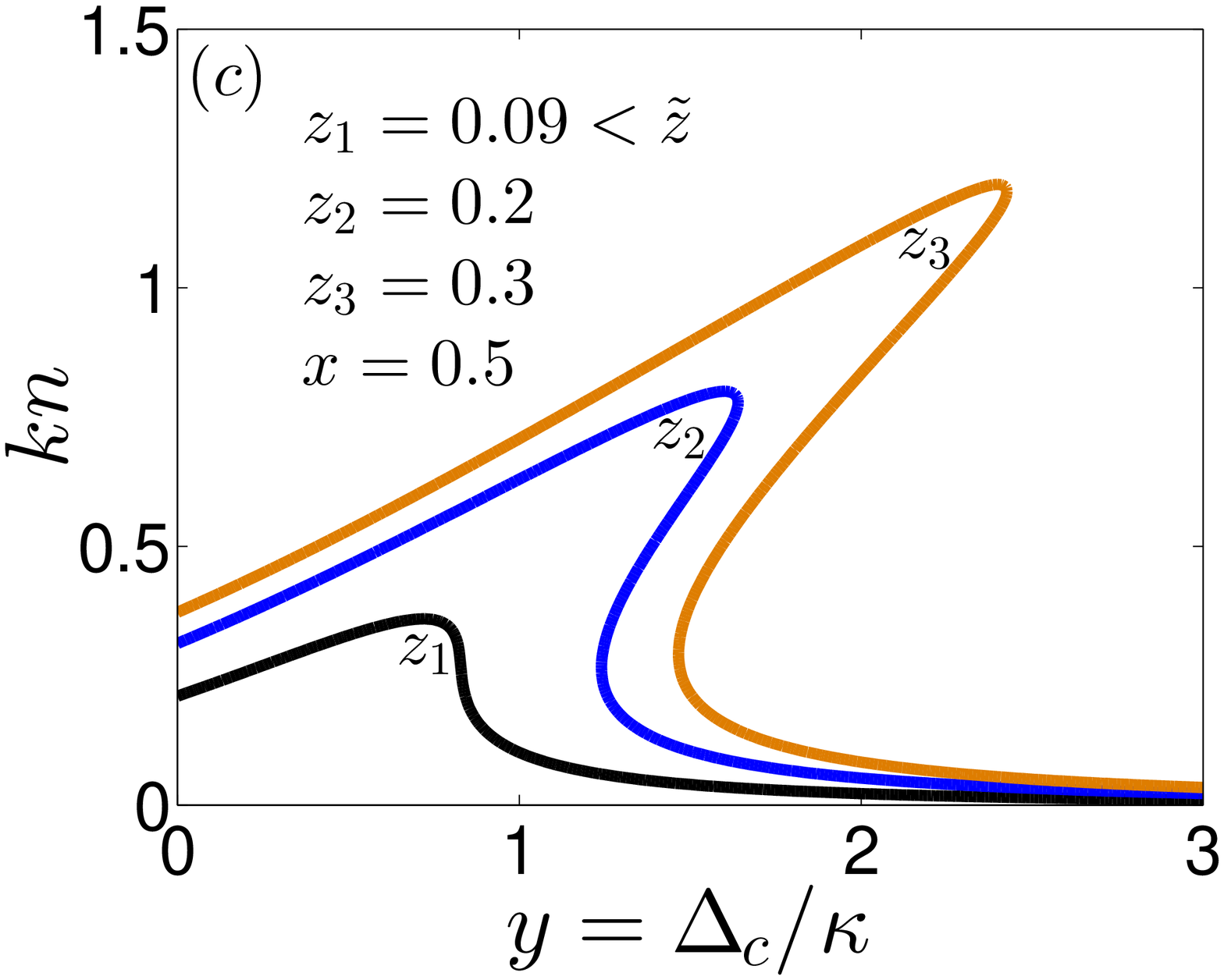}\includegraphics[bb=0 192 570 632,width=0.25\textwidth, clip]{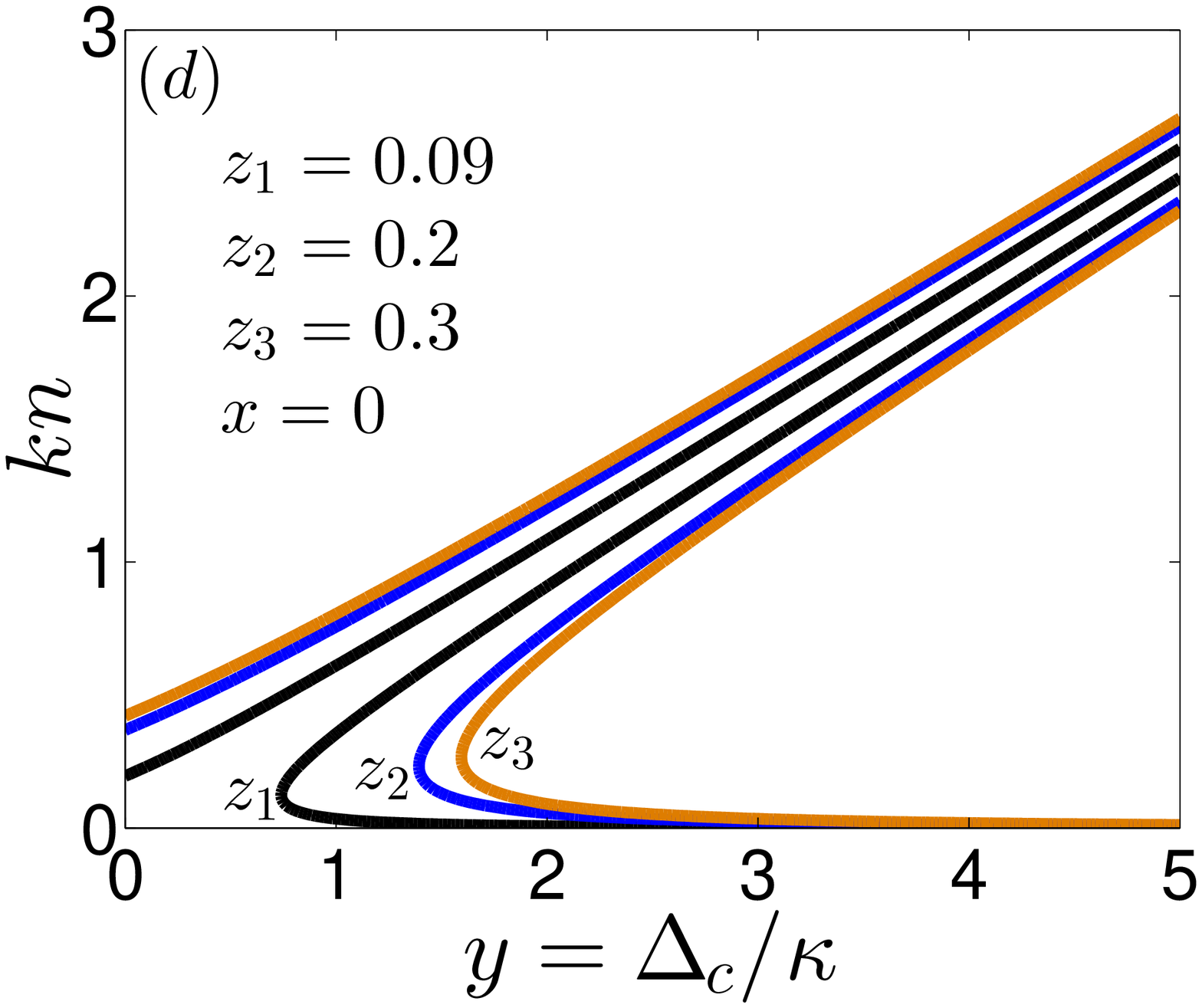}\newline%
\includegraphics[bb=0 192 570 632,width=0.25\textwidth, clip]{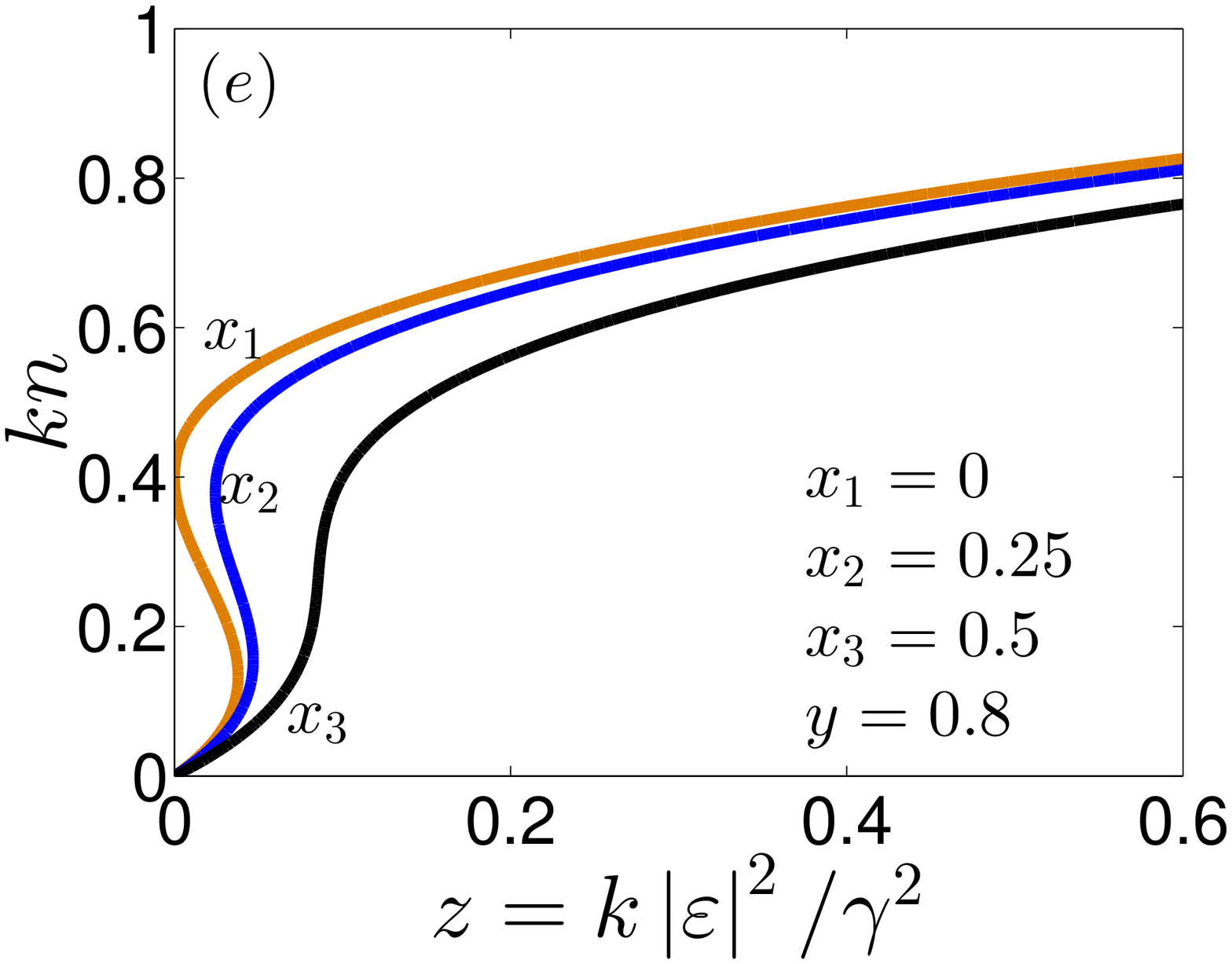}\includegraphics[bb=0 192 570 632,width=0.25\textwidth, clip]{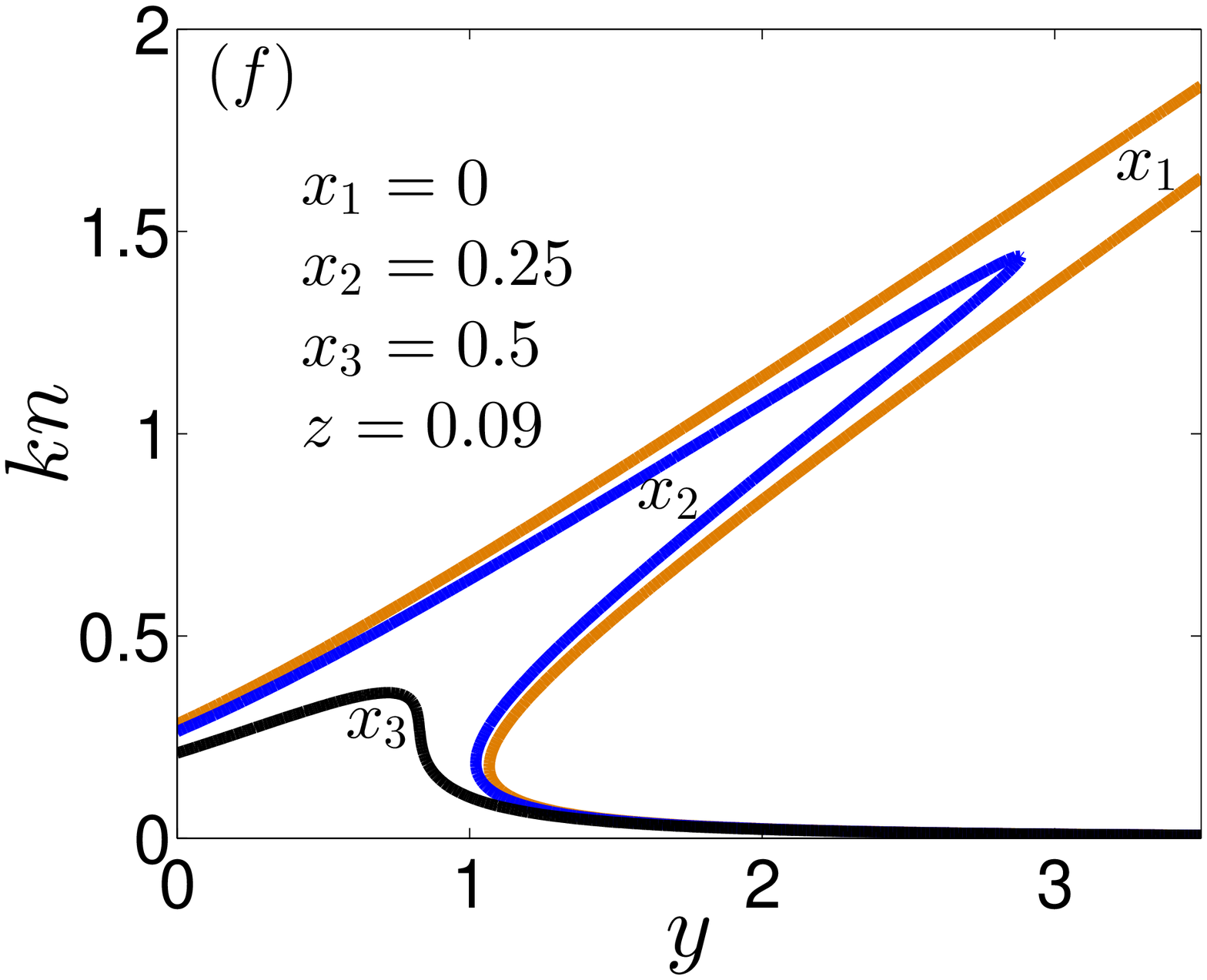}\newline%
\caption{(Color online) Optical bistability in the semiclassical regime.
Typical curves for the mean-field cavity occupation $kn$ as a function of the
dimensionless driving power $z$ (a,b), the detuning parameter $y$ (c,d) and
the detuning parameter $x$ (e,f). We show the mean-field occupation $kn$ as a
function of driving power $z$ for fixed detuning $y$ and the detuning
parameter$\ x$ of cavity A was set to $0.5$ in (a) and $0$ in (b). We show the
mean-field occupation $kn$ as a function of detuning $y$ for fixed driving
power $z$. The detuning parameter$\ x$ was set to $0.5$ in (c) and $0$ in (d).
We also show the mean-field occupation $kn$ as a function of driving power $z$
for fixed detuning $y$ and detuning parameter $x_{1}=0,x_{2}=0.25,x_{3}=0.5$
in (e), and $kn$ as a function of detuning parameter $y$ for fixed driving
power $z$ and detuning parameter $x_{1}=0,x_{2}=0.25,x_{3}=0.5$ in (f).}%
\label{fig4}%
\end{figure}

Using Eqs.~(\ref{A0}-\ref{C0}), we can obtain a third-order polynomial root
equation for the normalized mean photon number in the cavity of the
optomechanical system
\begin{equation}
f(\lambda)=4\lambda^{3}-4y\lambda^{2}+(x^{2}+y^{2})\lambda-z=0.\label{dxs}%
\end{equation}
Compare Eq.~(\ref{dxs}) with that in Ref.~\cite{80}, we find that
Eq.~(\ref{dxs}) has the same form as that in Ref.~\cite{80} when $x=0.5$.
Moreover, the coefficient $y$ has also been changed by the feedback control.
Here, we have used the condition $Q_{\mathrm{m}}=\omega_{\mathrm{m}}%
/\gamma_{\mathrm{m}}\gg1$ and introduced several dimensionless parameters
\begin{align}
\;\;\;x &  =\frac{p_{\mathrm{1}}}{\gamma},\;\;\;y=\frac{p_{\mathrm{2}}}%
{\gamma},\;\;\;z=k\frac{\left\vert \varepsilon\right\vert ^{\mathrm{2}}%
}{\gamma^{\mathrm{2}}},\label{param}\\
\;\;k &  =\frac{g_{_{\mathrm{0}}^{\mathrm{2}}}}{\gamma\omega_{\mathrm{m}}%
},\;\;\lambda=kn,\;\;\;n=\left\vert C_{\mathrm{0}}\right\vert ^{\mathrm{2}%
},\nonumber
\end{align}
with
\begin{align}
p_{\mathrm{2}} &  =\Delta_{\mathrm{c}}+\frac{4\gamma\sqrt{\kappa
\kappa_{\mathrm{f}}}\Delta_{\mathrm{s}}}{[4\Delta_{\mathrm{s}}^{\mathrm{2}%
}+(\sqrt{\kappa}+\sqrt{\kappa_{\mathrm{f}}})^{\mathrm{4}}]},\label{P1}\\
p_{\mathrm{1}} &  =\frac{\gamma}{2}-\frac{2\gamma\sqrt{\kappa\kappa
_{\mathrm{f}}}(\sqrt{\kappa}+\sqrt{\kappa_{\mathrm{f}}})^{\mathrm{2}}%
}{[4\Delta_{\mathrm{s}}^{\mathrm{2}}+(\sqrt{\kappa}+\sqrt{\kappa_{\mathrm{f}}%
})^{\mathrm{4}}]}.
\end{align}
It can be found that $k$ characterizes the optical nonlinearity induced by the
mechanical resonator, and the parameters $x$ and $y$ are determined by the
detuning frequencies $\Delta_{\mathrm{c}}$ and $\Delta_{\mathrm{s}}$ when all
of decay rates are fixed. $z$ is called the normalized power of the driving
field and $n$ is the mean photon number inside the cavity of the
optomechanical system.

We can also find that the mean photon number $n_{\mathrm{A}}=|A_{\mathrm{0}%
}|^{\mathrm{2}}$ of the controlled cavity can be given by
\begin{equation}
n_{\mathrm{A}}=Kn,
\end{equation}
with
\begin{equation}
K=\frac{4\gamma\kappa_{\mathrm{f}}}{4\Delta_{\mathrm{s}}^{\mathrm{2}}%
+(\sqrt{\kappa}+\sqrt{\kappa_{\mathrm{f}}})^{\mathrm{4}}},
\end{equation}
which shows that the mean photon number $n_{\mathrm{A}}$ of the controlled
cavity is proportional to the mean photon number $n$ of the controller cavity.
Without loss of generality, we assume that $K=1$ for the following discussions
such that both $n_{\mathrm{A}}$ and $n$ simply satisfy Eq.~(\ref{dxs}), which
can have either one or three roots, depending on the dimensionless parameters
$x$, $y$ and $z$.

In fact, when Eq.~(\ref{dxs}) is solved, three real roots can be found only
if: (i) the parameters $y$ and $z$ are larger than threshold values $\tilde
{y}$ and $\tilde{z}$ for given parameter $x$, that is,
\begin{align}
y  &  >\tilde{y}=\sqrt{3}x,\label{yuzhi}\\
z  &  >\tilde{z}=\frac{4}{27}\tilde{y}^{\mathrm{3}}, \label{yuzhiz}%
\end{align}
and (ii) the parameter $z$ must be in the region $z_{\mathrm{-}}%
(y)<z<z_{\mathrm{+}}(y)$ with
\begin{equation}
z_{\mathrm{\pm}}(y)=\frac{1}{27}[y(y^{\mathrm{2}}+3\tilde{y}^{\mathrm{2}}%
)\pm(y^{\mathrm{2}}-\tilde{y}^{\mathrm{2}})^{3/2}].
\end{equation}

The first externally controllable parameter, that can be used to control the
bistability behavior, is the normalized driving power $z$. In Fig.~\ref{fig4}%
(a), we show the normalized mean photon number $kn$ as a function of $z$ for
$x=0.5$, corresponding to the threshold value $\tilde{y}=\sqrt{3}/2$, and
three different parameters $y_{\mathrm{1}}=0.8<\sqrt{3}/2$, $y_{\mathrm{2}%
}=1.2>\sqrt{3}/2$ and$\;y_{\mathrm{3}}=1.7>\sqrt{3}/2$, respectively.
Fig.~\ref{fig4}(a) clearly shows that the optical bistability cannot be
observed when $y_{\mathrm{1}}<\sqrt{3}/2$ in contrast that the bistability can
be observed when $y_{\mathrm{2}},y_{\mathrm{3}}>\sqrt{3}/2$. From
Eq.~(\ref{yuzhi}) and Eq.~(\ref{P1}), we can find that the threshold value
$\tilde{y}$ is originated from the parameter $x$ which can be tuned from $0$
to $0.5$ by changing the frequency detuning $\Delta_{\mathrm{s}}$. In
Fig.\ref{fig4}(b), we show $kn$ as a function of $z$ for $x=0$, corresponding
to the threshold value $\tilde{y}=0$, and three different values of $y$ as
before. Under this condition, we find that the optical bistability phenomenon
can be observed with the increase of $z$ for arbitrary value of $y$ with $y>0$.

The second important externally controllable parameter, which can be used to
control the bistability behavior, is $y$ when $x$ is given. In Fig.~\ref{fig4}%
(c) we plot the normalized mean-photon number $kn$ as a function of detuning
$y$ for $x=0.5$, corresponding to the threshold value $\tilde{z}=1/6\sqrt{3}$,
and three different values $z_{\mathrm{1}}=0.09<1/6\sqrt{3}=\tilde{z}$,
$z_{\mathrm{2}}=0.2>1/6\sqrt{3}$, and $z_{\mathrm{3}}=0.3>1/6\sqrt{3}$,
respectively. Fig.~\ref{fig4}(c) shows that the optical bistability phenomenon
cannot be observed for $z<\tilde{z}$ in contrast to the case $z>\tilde{z}$.
Eq.~(\ref{yuzhi}) and Eq.~(\ref{yuzhiz}) show that the threshold value
$\tilde{z}$ is proportional to the detuning parameter $x$. This means that the
threshold value $\tilde{z}$ can be reduced to zero when $x=0$. In
Fig.~\ref{fig4}(d), the normalized photon number $kn$ is plotted as a function
of $y$ for $x=0$, corresponding to the threshold value $\tilde{z}=0$ of $z$,
and three different parameters $z_{\mathrm{1}}$, $z_{\mathrm{2}}$ and
$z_{\mathrm{3}}$ of the parameter $z$. Fig.~\ref{fig4}(d) shows that the
optical bistability phenomenon can be observed with the increase of $y$ for
arbitrary value of $z$ with $z>0$.

We now study how the parameter $x$ affects the bistability. In Fig.~\ref{fig4}%
(e), we plot $kn$ as a function of $z$ for $y=0.8$ and three different
parameters $x_{\mathrm{1}}=0$, $x_{\mathrm{2}}=0.25$ and $x_{\mathrm{3}}=0.5$.
The parameter $x_{\mathrm{1}}=0$ corresponds to the zero threshold value for
the parameters $y$ and $z$. The parameter $x_{\mathrm{2}}=0.25$ corresponds to
the threshold values $\tilde{y}=x_{2}\sqrt{3}$ and $\tilde{z}=4\tilde
{y}^{\mathrm{3}}/27$ for the parameters $y$ and $z$ respectively. However, the
parameter $x_{\mathrm{3}}=0.5$ corresponds to the maximum threshold values of
the parameter $y$ and $z$. It is clear that $y=0.8$ is larger than the
threshold values $\tilde{y}=0$ corresponding to $x_{\mathrm{1}}$ and
$\tilde{y}=0.433$ corresponding to $x_{\mathrm{2}}$, but $y=0.8$ is blow the
threshold value $\tilde{y}=0.87$ corresponding to $x_{\mathrm{3}}$. Thus,
Fig.~\ref{fig4}(e) clearly shows that $kn$ as a function of $z$ exists
bistable behavior for the parameters $x_{\mathrm{1}}$ and $x_{\mathrm{2}}$,
but bistable behavior cannot be found for the given parameter $x_{\mathrm{3}}%
$. In Fig.~\ref{fig4}(f) we show $kn$ as a function of $y$ for given parameter
$z=0.09$ and three same values $x_{\mathrm{1}}=0$, $x_{\mathrm{2}}=0.25$ and
$x_{\mathrm{3}}=0.5$ of the parameter $x$ as in Fig.~\ref{fig4}(e). Similar to
Fig.~\ref{fig4}(e), we find that $kn$ curve exhibits bistable behavior for
$x_{\mathrm{1}}$, corresponding to $\tilde{z}=0$, and $x_{\mathrm{2}}$,
corresponding to $\tilde{z}=0.012$. However, we cannot find bistable behavior
for $kn$ curve with the parameter $x_{\mathrm{3}}$ corresponding to $\tilde
{z}=0.096$.

According to above discussions, we can conclude that the nonlinear behavior in
the optomechanical system can be transferred to the linear cavity resonator
and controlled by using coherent feedback loop. The controlled bistability
phenomena can be realized by adjusting the power and the frequency of
externally applied driving field. In contrast to the bistability of
optomechanical system~\cite{80} and by comparing Fig.~\ref{fig4}(a) with
Fig.~\ref{fig4}(b), or comparing Fig.~\ref{fig4}(c) with Fig.~\ref{fig4}(d),
we find that the threshold value of the bistable behavior can be reduced to
zero even in the optomechanical system when the coherent feedback control is introduced.

\section{COHERENT FEEDBACK INDUCED PHOTON BLOCKADE}

We now turn to consider the case that the external driving field is weak
enough and both cavity fields are in the quantum regime. We will study the
single-photon blockade effect in the linear cavity induced by the coherent
feedback control using an optomechanical system as a controller. The evolution
of the coherent-feedback controlled system, comprised of the linear cavity and
the optomechanical system, is described by the Hamiltonian of Eq.~(\ref{Htot}%
). In the rotating reference frame at the frequency $\omega_{d}$ of the
driving field and under the Born-Markov approximation, the time evolution of
the whole system is described by a Lindblad-type master equation~\cite{81,82}
\begin{equation}
\frac{d\rho}{dt}=\mathcal{L}(\rho)=-i[\tilde{H},\rho]+\frac{1}{2}(2\tilde
{L}\rho\tilde{L}^{\dag}-\tilde{L}^{\dag}\tilde{L}\rho-\rho\tilde{L}^{\dag
}\tilde{L})+\gamma_{\mathrm{m}}\mathcal{D}(b), \label{master}%
\end{equation}
with
\begin{align}
\tilde{L}  &  =(\sqrt{\kappa}+\sqrt{\kappa_{\mathrm{f}}})a+\sqrt{\gamma}c,\\
\mathcal{D}(b)  &  =b\rho b^{\dag}-\frac{1}{2}\rho b^{\dag}b-\frac{1}%
{2}b^{\dag}b\rho.
\end{align}
Here, we assume that the mechanical resonator and the cavity fields are in the
zero-temperature environment for the convenience of calculations. The
steady-state density operator $\rho_{ss}$ can be obtained by setting
$d\rho/dt=\mathcal{L}(\rho)=0$, and then the normalized equal-time
second-order correlation functions for both cavity fields in the steady-state
case can be calculated by
\begin{align}
g_{\mathrm{a}}^{\mathrm{(2)}}(0)  &  =\frac{\left\langle a^{\dag}a^{\dag
}aa\right\rangle }{\left\langle a^{\dag}a\right\rangle ^{\mathrm{2}}}%
=\frac{\mathrm{Tr}(\rho_{\mathrm{ss}}a^{\dag}a^{\dag}aa)}{[\mathrm{Tr}%
(\rho_{\mathrm{ss}}a^{\dag}a)]^{\mathrm{2}}},\label{second}\\
g_{\mathrm{c}}^{\mathrm{(2)}}(0)  &  =\frac{\left\langle c^{\dag}c^{\dag
}cc\right\rangle }{\left\langle c^{\dag}c\right\rangle ^{\mathrm{2}}}%
=\frac{\mathrm{Tr}(\rho_{\mathrm{ss}}c^{\dag}c^{\dag}cc)}{[\mathrm{Tr}%
(\rho_{\mathrm{ss}}c^{\dag}c)]^{\mathrm{2}}}.
\end{align}
Here, the subscripts $a$ and $c$ denote the cavity fields inside the
controlled cavity and the cavity of the optomechanical system. Below, we will
study the single-photon blockade effect in the linear cavity using
Eq.~(\ref{second}). The single-photon blockade effect in the cavity of the
optomechanical system can also be studied in a similar way.

\subsection{Photon blockade in optomechanical single-photon strong-coupling
regime}

\begin{figure}[h]
\includegraphics[bb=8 190 553 634,  width=0.23\textwidth, clip]{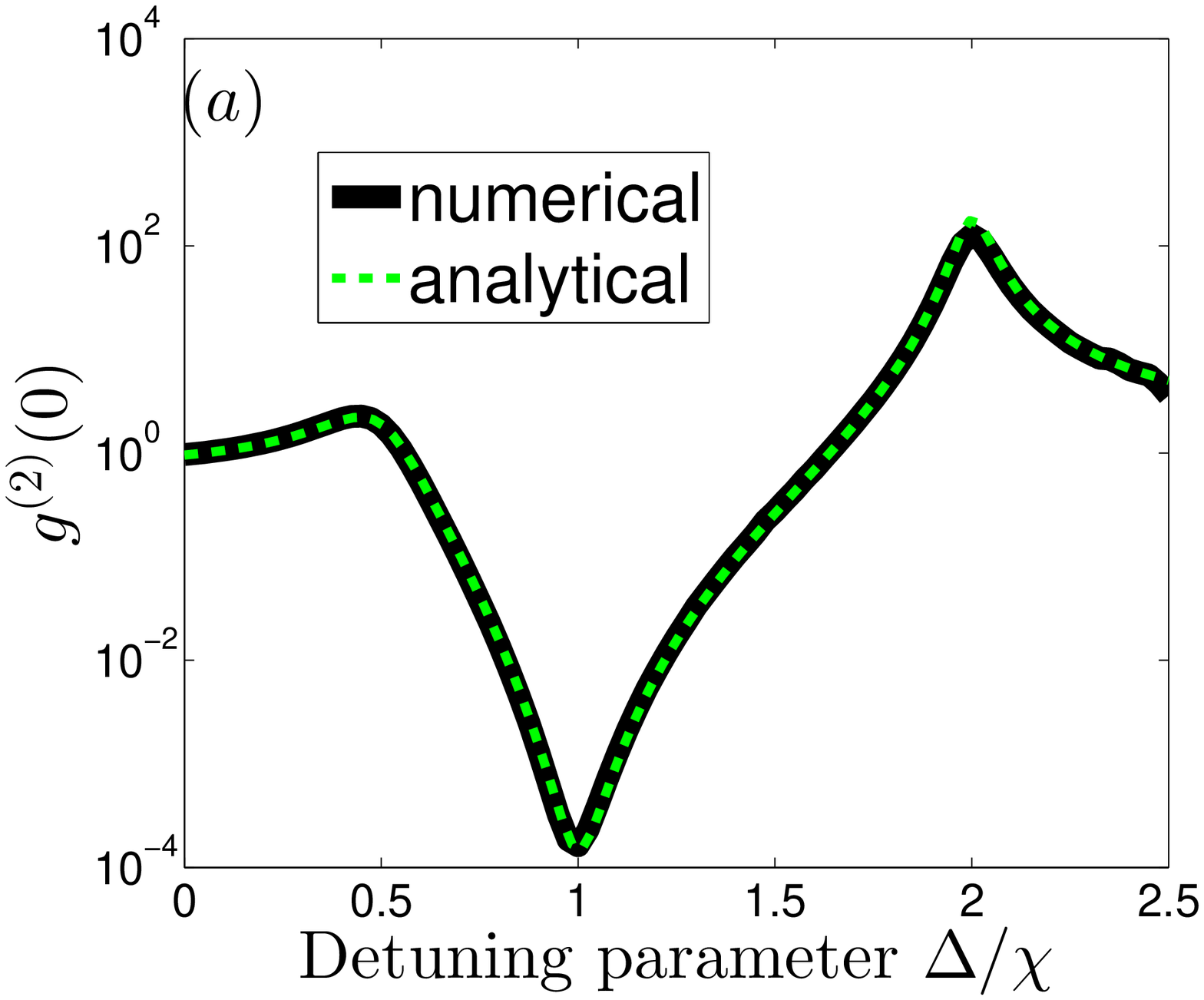}
\includegraphics[bb=8 190 553 634,  width=0.23\textwidth, clip]{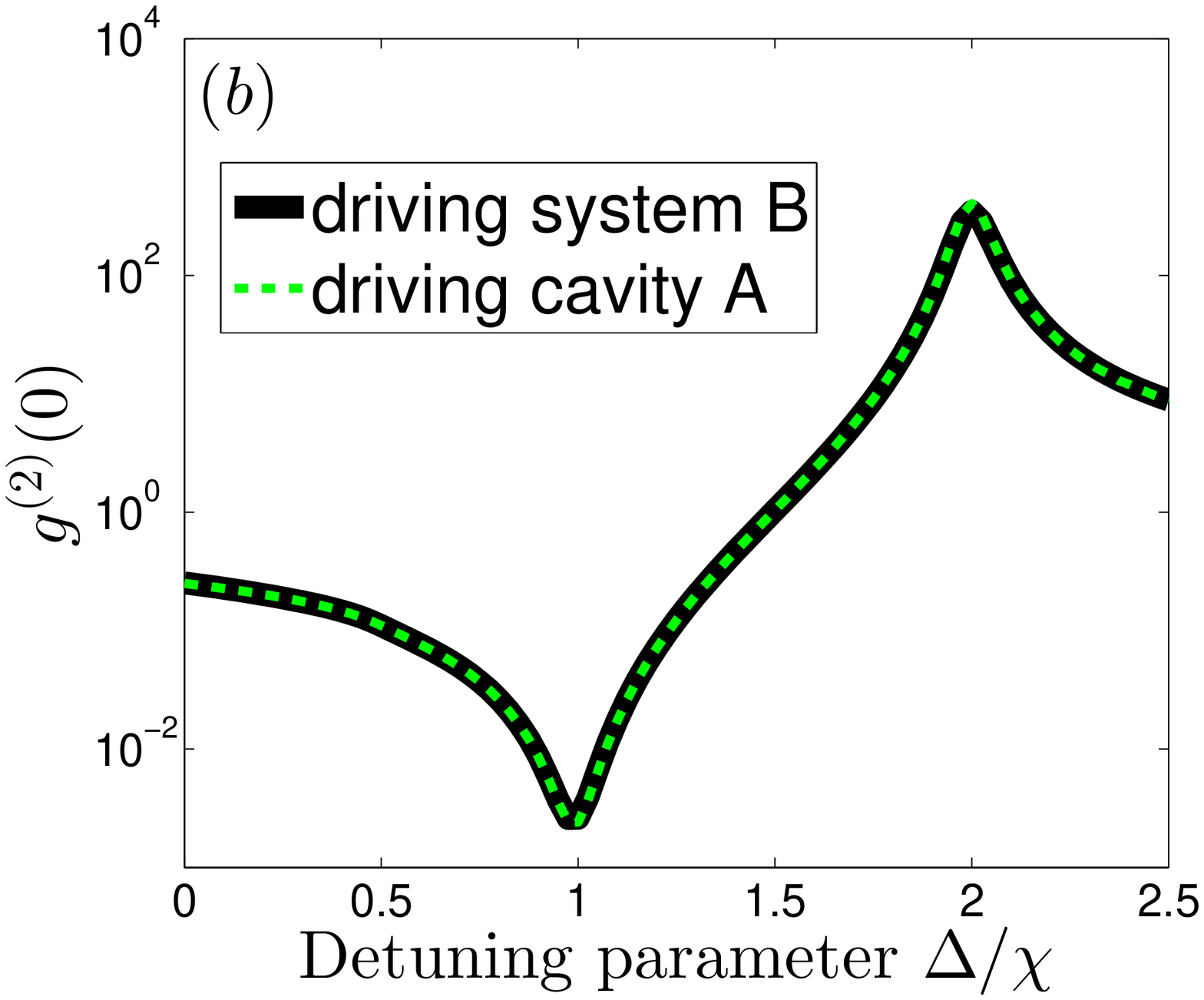}
\caption{Color online) The numerical and analytical solutions for second-order
correlation function$g_{\mathrm{a}}^{\mathrm{(2)}}(0)$ of linear cavity field
versus detuning parameter $\Delta/\chi$ are shown in (a). The second-order
correlation functions $g_{\mathrm{c}}^{\mathrm{(2)}}(0)$ of optomechanical
cavity field versus detuning parameter $\Delta/\chi$ by driving controlled
cavity or controller cavity are shown in (b). The system parameters for this
simulation are: $\kappa=\gamma$, $\kappa_{\mathrm{f}}=\gamma$, $g_{\mathrm{0}%
}=32\gamma$, $\omega_{\mathrm{m}}=100\gamma$, $\gamma_{\mathrm{m}}=0.01\gamma
$, $\epsilon=0.1\gamma$, $\gamma/2\pi=1$ MHz.}%
\label{fig5}%
\end{figure}

We now consider the statistical properties of photons when the optomechanical
system is assumed to be in single-photon strong-coupling regime. For
comparison, the second-order correlation functions $g_{\mathrm{a}%
}^{\mathrm{(2)}}(0)$ calculated by master equation and analytical solution are
shown in Fig.~\ref{fig5}. To analytically give the condition for photon
blockade in the controlled cavity, the solution of $g_{\mathrm{a}%
}^{\mathrm{(2)}}(0)$ is obtained by the following method. With definition of
the Kerr nonlinear coefficient $\chi=g_{\mathrm{0}}^{\mathrm{2}}%
/\omega_{\mathrm{m}}$, in the rotating reference frame at the frequency
$\omega_{\mathrm{d}}$ of the driving field and under the condition
$g_{_{\mathrm{0}}}$/$\omega_{\mathrm{m}}\ll1$, the effective Hamiltonian in
Eq.~(\ref{Htot}) can be written as
\begin{align}
\tilde{H}_{\mathrm{eff}}  &  =\Delta_{\mathrm{s}}a^{\dag}a+(\Delta
_{\mathrm{c}}-\chi)c^{\dag}c-\chi c^{\dag}cc^{\dag}c+\epsilon(c^{\dag
}+c)\nonumber\\
&  +\frac{i}{2}(\sqrt{\gamma\kappa}-\sqrt{\gamma\kappa_{\mathrm{f}}})(a^{\dag
}c-c^{\dag}a)+\omega_{\mathrm{m}}b^{\dag}b. \label{36}%
\end{align}
Eq.~(\ref{36}) shows that the mechanical mode can be decoupled from the
optical modes when all terms of parameter$\ g_{_{\mathrm{0}}}$/$\omega
_{\mathrm{m}}$\ are neglected, thus let us assume that the state of the whole
system is $\left\vert \psi\right\rangle =\left\vert \varphi\right\rangle
\left\vert \phi\right\rangle _{\mathrm{m}}$ where $\left\vert \varphi
\right\rangle $ denotes the photon states of both cavity fields and
$\left\vert \phi\right\rangle _{\mathrm{m}}$ denotes the phonon states. Under
the weak-driving condition, i.e. $\epsilon/\gamma\rightarrow0$, we can
describe the photon states as
\begin{equation}
\left\vert \varphi\right\rangle =%
{\displaystyle\sum\limits_{n_{\mathrm{a}}=0}^{2}}
{\displaystyle\sum\limits_{n_{\mathrm{c}}=0}^{2}}
C_{n_{\mathrm{a}},n_{\mathrm{c}}}\left\vert n_{\mathrm{a}},n_{\mathrm{c}%
}\right\rangle ,
\end{equation}
where $\left\vert n_{\mathrm{a}}\right\rangle $ and $\left\vert n_{\mathrm{c}%
}\right\rangle $ are the photon states of the controlled cavity and the cavity
of optomechanical controller, respectively. It is obvious that $C_{\mathrm{00}%
}\gg C_{\mathrm{01}},C_{\mathrm{10}}\gg C_{\mathrm{11}},C_{\mathrm{20}%
},C_{\mathrm{02}}$ under the weak driving condition $\epsilon/\gamma
\rightarrow0$. From Eq.~(\ref{second}), we can find
\begin{equation}
g_{\mathrm{a}}^{\mathrm{(2)}}(0)=\frac{\mathrm{Tr}(\rho_{\mathrm{ss}}a^{\dag
}a^{\dag}aa)}{[\mathrm{Tr}(\rho_{\mathrm{ss}}a^{\dag}a)]^{\mathrm{2}}}=\frac{%
{\displaystyle\sum\limits_{n_{\mathrm{a}}}}
n_{\mathrm{a}}(n_{\mathrm{a}}-1)P_{n_{\mathrm{a}}}}{[%
{\displaystyle\sum\limits_{n_{\mathrm{a}}}}
n_{\mathrm{a}}P_{n_{\mathrm{a}}}]^{\mathrm{2}}}.
\end{equation}
in the steady-state. Here, $P_{n_{\mathrm{a}}}$ represents the probability of
$n_{\mathrm{a}}$ photons distribution which can be expressed as%
\begin{equation}
P_{n_{\mathrm{a}}}=%
{\displaystyle\sum\limits_{n_{\mathrm{c}}}}
\left\vert C_{n_{\mathrm{a}},n_{\mathrm{c}}}\right\vert ^{2}.
\end{equation}
which means $P_{n_{\mathrm{a}}}\gg P_{n_{\mathrm{a}}+1}$ for $n_{\mathrm{a}%
}\geq2$.

To obtain the steady-state solution, we phenomenologically introduce the
non-Hermitian complex Hamiltonian by setting the parameters as
\begin{align}
\Delta_{\mathrm{s}}  &  \longrightarrow\Delta_{\mathrm{s}}-i(\sqrt{\kappa
}+\sqrt{\kappa_{\mathrm{f}}})^{2}/2,\\
\Delta_{\mathrm{c}}  &  \longrightarrow\Delta_{\mathrm{c}}-i\gamma/2.
\end{align}
in the Hamiltonian~(\ref{36}). Thus, under weak driving limit that the
probabilities for more than three photon can be neglected, the second-order
correlation function $g_{\mathrm{a}}^{\mathrm{(2)}}(0)$ can be expressed
as~\cite{Z.P.L}%
\begin{equation}
g_{\mathrm{a}}^{\mathrm{(2)}}(0)=\frac{\left\vert \Delta_{\mathrm{s}}%
+\Delta_{\mathrm{c}}-2\chi-i\kappa_{\mathrm{a}}/2\right\vert ^{\mathrm{2}%
}\left\vert \Delta_{\mathrm{c}}-\chi-i\gamma/2\right\vert ^{\mathrm{2}}%
}{\left\vert (\Delta_{\mathrm{s}}+\Delta_{\mathrm{c}}-\chi-i\kappa
_{{\normalsize a}}/2)(\Delta_{\mathrm{c}}-2\chi-i\gamma/2)\right\vert
^{\mathrm{2}}}.
\end{equation}
with
\begin{equation}
\kappa_{a}=(\sqrt{\kappa}+\sqrt{\kappa_{\mathrm{f}}})^{\mathrm{2}}+\gamma.
\end{equation}
Without loss of generality, we consider $\Delta_{\mathrm{s}}=\Delta
_{\mathrm{c}}=\Delta$ which means that the controller cavity and the
controlled cavity resonator are resonate with each other, thus the analytical
solution of $g_{\mathrm{a}}^{\mathrm{(2)}}(0)$ is simplified to%
\begin{equation}
g_{\mathrm{a}}^{\mathrm{(2)}}(0)=\frac{[4(\Delta-\chi)^{\mathrm{2}}%
+\kappa_{\mathrm{a}}^{\mathrm{2}}/4][(\Delta-\chi)^{\mathrm{2}}+\gamma
^{\mathrm{2}}/4]}{\left\vert (2\Delta-\chi-i\kappa_{\mathrm{a}}/2)(\Delta
-2\chi-i\gamma/2)\right\vert ^{\mathrm{2}}}. \label{G2aa}%
\end{equation}

From Eq.~(\ref{G2aa}) we can find that when the decay rates
$\gamma$ and $\kappa_{\mathrm{a}}$ are far less than the cavity detuning
frequency $\Delta$\ and the Kerr nonlinear coefficient $\chi$, the
second-order correlation function $g_{\mathrm{a}}^{\mathrm{(2)}}(0)$ becomes
far less than one at the point of $\Delta/\chi=1$. This means that the photon
blockade occurs. In the following, we numerically study two very different
parameter regions: (i) $\kappa=\kappa_{\mathrm{f}}=\gamma$ and $g\sim
\omega_{\mathrm{m}}$ which are not easy to be achieved for current
experiments; (ii) $\kappa=\kappa_{\mathrm{f}}\gg\gamma$ and $g\ll
\omega_{\mathrm{m}}$ which are possible experimentally.

The second-order correlation function $g_{\mathrm{a}}^{\mathrm{(2)}}(0)$,
calculated by master equation and analytical solution, as a function of
$\Delta/\chi$ is shown in Fig.~\ref{fig5}(a) for $\kappa=\kappa_{\mathrm{f}%
}=\gamma$ and $g_{\mathrm{0}}\sim\omega_{\mathrm{m}}/3$. We find that the
results obtained by numerical calculations and approximated solutions are
almost same. Figure~\ref{fig5}(a) shows that there is a minimum value at
$\Delta/\chi=1$ with $g_{\mathrm{a}}^{\mathrm{(2)}}(0)\ll1$, which means that
the photon blockade occurs and the single-photon can come out of the
controlled cavity one by one. Figure~\ref{fig5}(a) also shows that there is a
maximum value at $\Delta/\chi=2$ with $g_{\mathrm{a}}^{\mathrm{(2)}}(0)\gg1$,
which means that the two-photon tunneling occurs and photons can come out of
the controlled cavity in pairs. These characteristics have been
shown~\cite{64,65,83} in the standard optomechanical systems. However, the
strong photon blockade and two-photon tunneling studied here are found in the
controlled linear cavity, which is due to the coherent-feedback through the
nonlinear quantum controller. This can be verified through $g_{\mathrm{c}%
}^{\mathrm{(2)}}(0)$, shown in Fig.~\ref{fig5}(b), of the cavity field of the
optomechanical system. We find a minimum value at $\Delta/\chi=1$
corresponding to $g_{\mathrm{c}}^{\mathrm{(2)}}(0)\ll1$ and a maximum value at
$\Delta/\chi=2$ corresponding to $g_{\mathrm{c}}^{\mathrm{(2)}}(0)\gg1$. Thus,
we would like to say that the photon blockade in the controller cavity is
transferred to the controlled cavity by the coherent feedback loop. To answer
the question wether the different driving strategy, such as driving the
controlled cavity in contrast to driving controller cavity, can result in
different optical nonlinear behavior, $g_{\mathrm{c}}^{\mathrm{(2)}}(0)$ for
driving the controlled cavity is numerically studied and shown in
Fig.~\ref{fig5}(b), we find that there is no significant difference under
these two different driving ways by given other parameters.

\begin{figure}[ptb]
\includegraphics[bb=8 190 553 634,  width=0.23\textwidth, clip]{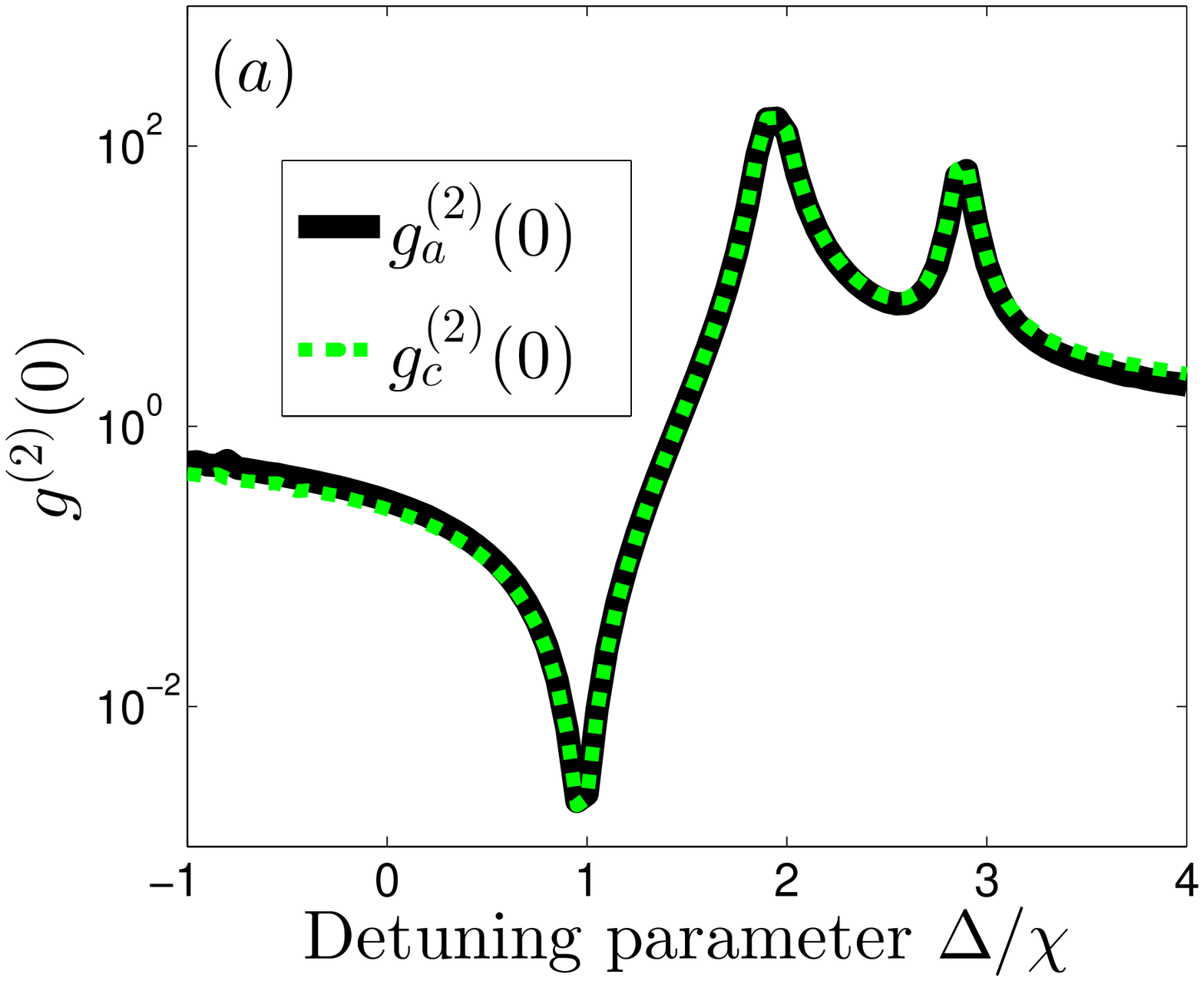}
\includegraphics[bb=8 190 553 634,  width=0.23\textwidth, clip]{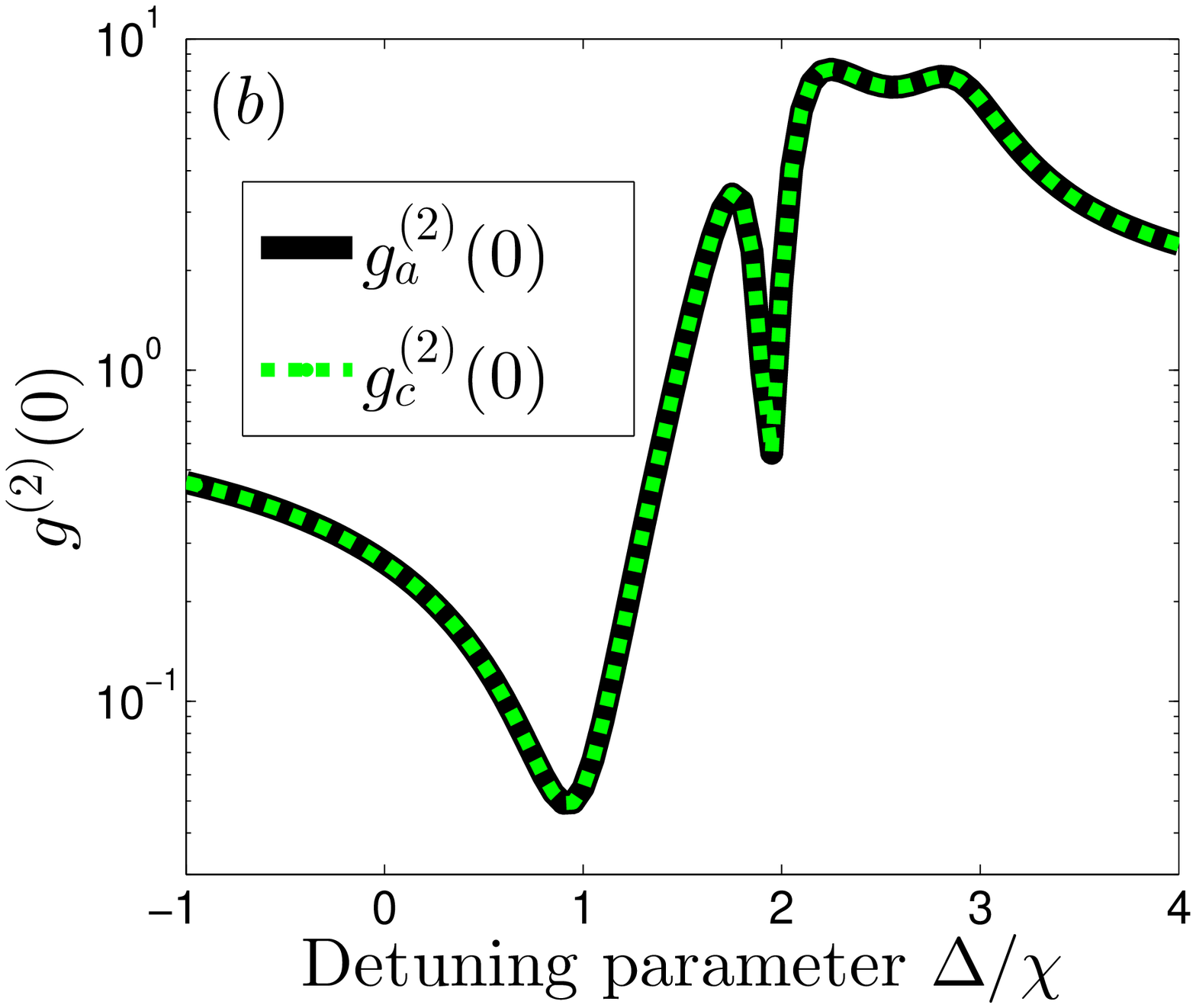}
\caption{(Color online) The second-order correlation functions $g_{\mathrm{a}%
}^{\mathrm{(2)}}(0)$ of controlled cavity field and $g_{\mathrm{c}%
}^{\mathrm{(2)}}(0)$ of optomechanical cavity field versus detuning parameter
$\Delta/\chi$ by driving the controlled cavity in (a) or driving the
controller cavity in (b). The system parameters for this simulation are:
$\kappa=10\gamma$, $\kappa_{\mathrm{f}}=10\gamma$, $g_{\mathrm{0}}=2.5\gamma$,
$\omega_{\mathrm{m}}=100\gamma$, $\gamma_{\mathrm{m}}=0.01\gamma$,
$\epsilon=0.01\gamma$, $\gamma/2\pi=1$ MHz.}%
\label{fig7}%
\end{figure}

Let us now study another parameter regions, e.g., $\kappa=\kappa_{\mathrm{f}%
}=10\gamma$ and $g_{\mathrm{0}}=\omega_{\mathrm{m}}/40$. The second-order
correlation functions of controlled cavity field and controller cavity field
versus $\Delta/\chi$ by driving the controlled cavity are shown in
Fig.~\ref{fig7}(a). We find that there exists a local minimum at $\Delta
/\chi=1$, a local maximum at $\Delta/\chi=2$ and another local maximum at
$\Delta/\chi=3$ for each curve. The minimum value corresponding to
$g_{\mathrm{a}}^{\mathrm{(2)}}(0)\ll1$ and $g_{\mathrm{c}}^{\mathrm{(2)}%
}(0)\ll1$ means that the photons in the controlled cavity and controller
cavity can be blockaded. The first local maximum at $\Delta/\chi=2$
corresponding to $g_{\mathrm{a}}^{\mathrm{(2)}}(0)\gg1$ and $g_{\mathrm{c}%
}^{\mathrm{(2)}}(0)\gg1$ means that the single-photon transition from ground
state to the first excited state is suppressed and second photon can enter the
driven cavity making resonant transition from ground state to second excited
state with the first photon. The second local maximum value at $\Delta/\chi=3$
corresponding to $g_{\mathrm{a}}^{\mathrm{(2)}}(0)\gg1$ and $g_{\mathrm{c}%
}^{\mathrm{(2)}}(0)\gg1$ means that the three-photon resonant excitation
happens and three-photon tunneling can be observed.

The second-order correlation functions of controlled cavity field and
controller cavity field versus $\Delta/\chi$ by driving the controller cavity
are shown in Fig.~\ref{fig7}(b). In contrast to the case that the controlled
cavity is driven, we find that there exist a global minimum point at
$\Delta/\chi=1$, a local minimum at $\Delta/\chi=2$ and local maximum at
$\Delta/\chi=3$ for each curve. The global minimum corresponding to
$g_{\mathrm{a}}^{\mathrm{(2)}}(0)\ll1$ and $g_{\mathrm{c}}^{\mathrm{(2)}%
}(0)\ll1$ means that the photons in controlled cavity and controller cavity
can be blockaded. The local minimum at $\Delta/\chi=2$ corresponding to
$g_{\mathrm{a}}^{\mathrm{(2)}}(0)<1$ and $g_{\mathrm{c}}^{\mathrm{(2)}}(0)<1$
means that the two-photon tunneling is suppressed and photon blockade happens.
The local maximum at $\Delta/\chi=3$ corresponding to $g_{\mathrm{a}%
}^{\mathrm{(2)}}(0)\gg1$ and $g_{\mathrm{c}}^{\mathrm{(2)}}(0)\gg1$ means that
the three-photon tunneling can be observed.

Comparing Fig.~\ref{fig7}(a) with Fig.~\ref{fig7}(b), and also in contrast to
Fig.~\ref{fig5}, we find that the second-order correlation functions present
different behaviors for the different driving strategies in the region around
the point $\Delta/\chi=2$ under the condition $\kappa=\kappa_{\mathrm{a}}%
\gg\gamma$. This difference comes from the unbalanced input-output rates of
controlled cavity and controller cavity. When the driving field is applied to
the controlled cavity and under the two-photon resonant driving($\Delta
/\chi=2$), due to the output rate $\kappa$ is much larger than the input rate
$\gamma$, the first photon in controlled cavity transports to the controller
cavity rapidly. Simultaneously, due to the output rates $\gamma$ is much less
than the input rates $\kappa_{\mathrm{f}}$, the second photon has entered into
the the controlled cavity before the first photon comes back. Therefore, at
last the two-photon tunneling can be observed when the second photon meets the
first one. However, When the driving field is applied to the controller cavity
and under the two-photon resonant driving($\Delta/\chi=2$), due to the output
rate $\gamma$ is much less than the input rate $\kappa_{\mathrm{f}}$, the
first photon entering controller cavity cannot be transported to controlled
cavity before the second photon is coming. Cooperated with the optomechanical
nonlinearity, the first photon in the controller cavity prevents the second
one from entering cavity and then the photon blockade occurs.

Comparing Fig.~\ref{fig5} with Fig.~\ref{fig7}, we would like to mention that
single-photon blockade can be more easily observed when the coherent-coupling
strength $\sqrt{\kappa\gamma}$ and $\sqrt{\kappa_{\mathrm{f}}\gamma}$ are
enhanced no matter the driving is applied to controlled cavity or controller
cavity. This is equivalent to enhance the coupling strength between two
cavities for controlled system and optomechanical controller~\cite{68}.

\subsection{Photon blockade in optomechanical single-photon weak-coupling
regime}

\begin{figure}[h]
\includegraphics[bb=8 190 553 634,  width=0.23\textwidth, clip]{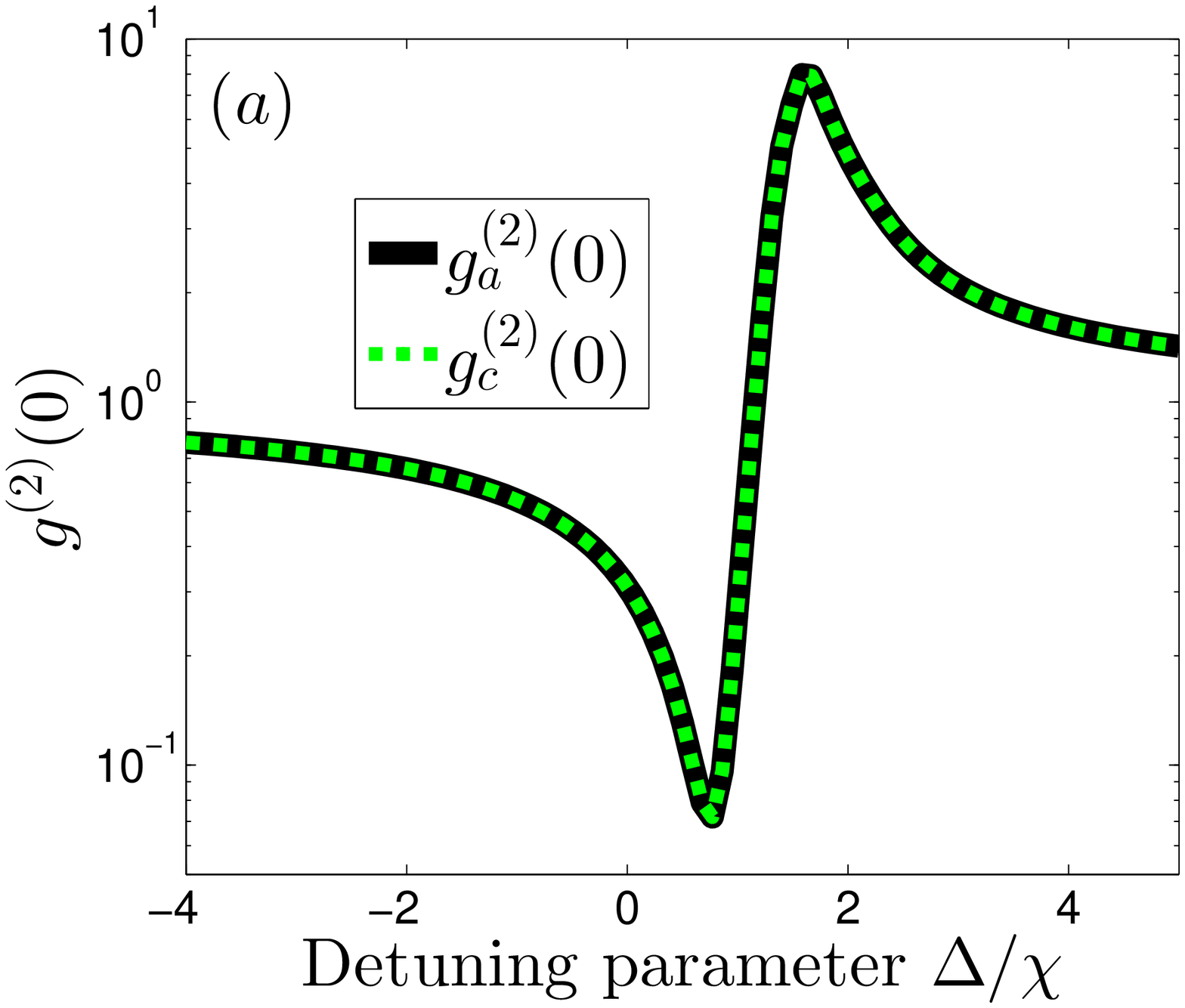}
\includegraphics[bb=8 190 553 634,  width=0.23\textwidth, clip]{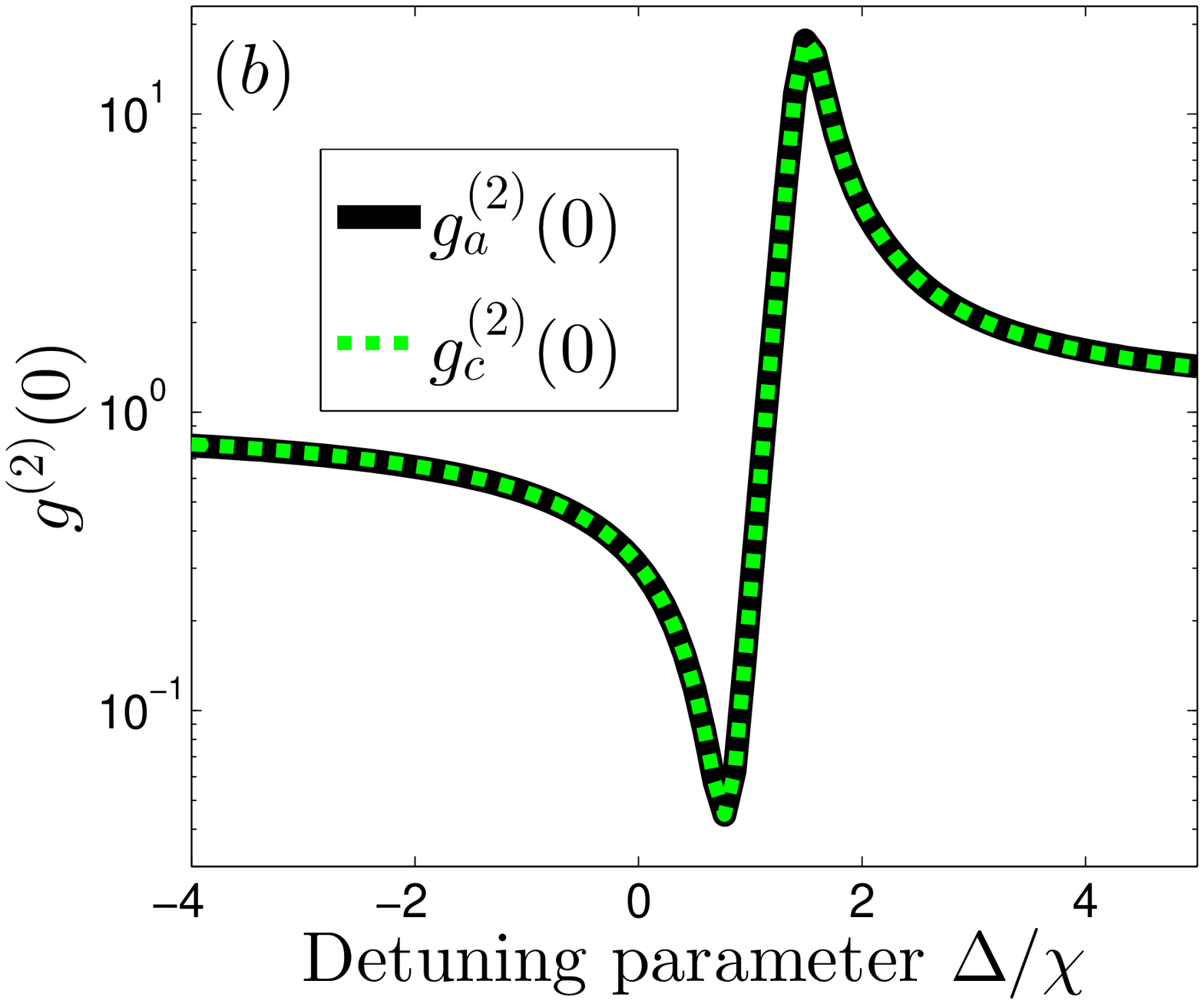}
\caption{(Color online) The second-order correlation functions $g_{a}%
^{(2)}(0)$ of linear cavity resonator and $g_{c}^{(2)}(0)$ of optomechanical
controller versus detuning parameter $\Delta/\chi$ by driving the
optomechanical controller in (a) or driving the controller cavity in (b). The
system parameters for this simulation are: $\kappa=\gamma$, $\kappa
_{f}=1.2\gamma$, $g_{0}=0.3\gamma$, $\omega_{m}=10\gamma$. The other
parameters are the same as in Fig.~\ref{fig7}.}%
\label{fig9}%
\end{figure}

Let us now study the possibility on the photon blockade using coherent
feedback control strategy in a weak single-photon optomechanical coupling
regime, e.g., $g_{\mathrm{0}}=0.3\gamma$. We focus on the statistical
properties of photons. The second-order correlation function of controlled
cavity field and controller cavity field versus $\Delta/\chi$ are studied when
the driving field is applied to the controller cavity.

As shown in Fig.~\ref{fig9}(a), we can find that there exists a minimum near
the point $\Delta/\chi=1$ corresponding to $g_{\mathrm{a}}^{\mathrm{(2)}%
}(0)\ll1$ and a maximum near the point $\Delta/\chi=2$ corresponding to
$g_{\mathrm{a}}^{\mathrm{(2)}}(0)\gg1$. This means that the photons in
controlled cavity can exhibit strong blockade and two-photon tunneling when it
is coherently feedback controlled by a optomechanical controller even in weak
optomechanical coupling. Fig.~\ref{fig9}(a) also clearly shows that the curve
of $g_{a}^{(2)}(0)$ fits well with the curve of $g_{c}^{(2)}(0)$ as well as
which has been shown in Fig.~\ref{fig7}. This is easily understood, because
the controller cavity and the controlled cavity are coherently coupled with
each other through the closed feedback loop with explicit photon flow
direction, which is naturally determined by the propagation of the quantum
field. Thus, if photon blockade (tunneling) occurs in the controller cavity,
then the photons from the controller cavity are output one by one (two by
two), and enter the controlled cavity, and vice versa.

\begin{figure}[h]
\includegraphics[bb=90 338 435 580,  width=0.35\textwidth, clip]{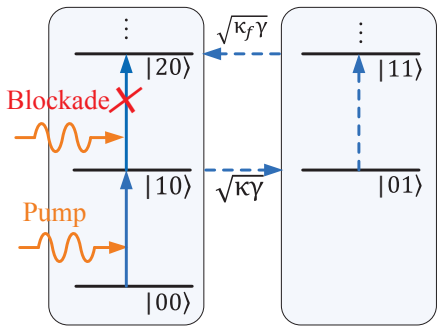}\caption{(Color
online) Schematic diagrams for two-photon exciting process. Two different
transition paths leading to the destructive quantum interference which is
responsible for the strong photon blockade. One path is the direct excitation
from $\left\vert 10\right\rangle $ to $\left\vert 20\right\rangle $ and the
other is drawn by the dotted arrows.}%
\label{fig12}%
\end{figure}

Similar to Fig.~\ref{fig9}(a), the second correlation functions $g_{\mathrm{a}%
}^{\mathrm{(2)}}(0)$ and $g_{\mathrm{c}}^{\mathrm{(2)}}(0)$ versus
$\Delta/\chi$ are shown in Fig.~\ref{fig9}(b) when the driving field is
applied to the controlled cavity. Similar to the case for driving the
controller cavity, a minimum near the point $\Delta/\chi=1$ corresponding to
$g_{\mathrm{a}}^{\mathrm{(2)}}(0),\,g_{\mathrm{c}}^{\mathrm{(2)}}(0)\ll1$ and
a maximum near the point $\Delta/\chi=2$ corresponding to $g_{\mathrm{a}%
}^{\mathrm{(2)}}(0),\,g_{\mathrm{c}}^{\mathrm{(2)}}(0)\gg1$ can be found. The
only difference is that the minimum (maximum) value for driving controlled
cavity is smaller (bigger) than that for driving controller cavity. That is,
the photon blockade (tunneling) is better in driving controlled cavity than
that for driving the controller cavity with the same parameters. We would also
like to mention that single-photon blockade can be more easily observed when
coherent-coupling strength $g_{\mathrm{0}}$ is enhanced no matter the driving
is applied to contolled cavity or controller cavity.

The strong photon blockade under weak optomechanical coupling comes from the
cooperation between the weak Kerr nonlinearity in controller cavity induced by
optomechanical interaction and the destructive interference for different
paths of two-photon excitation process~\cite{84,85}. In order to better
understand the interference process, two different paths for two-photon
excitations are shown in Fig.~\ref{fig12}. The first path is direct excitation
from one photon to two photons in the driven cavity($\left\vert
00\right\rangle \rightarrow\left\vert 20\right\rangle $). The second path is
one photon coherently passed to the other one and finally comes back to the
driven cavity($\left\vert 00\right\rangle \rightarrow\left\vert
10\right\rangle \rightarrow\left\vert 01\right\rangle \rightarrow\left\vert
11\right\rangle \rightarrow\left\vert 20\right\rangle $) which is
unidirectional and uniquely determined by the propagation direction of quantum
field. $\sqrt{\kappa\gamma}$ and $\sqrt{\kappa_{\mathrm{f}}\gamma}$ can be
considered as effective coupling strength between these two cavities, which is
the key to induce destructive interference. It is obviously that the cavities'
losses play a crucial role for achieving such destructive interference in
contrast to other photon blockade systems introduced so far in which the
cavities' losses always play a negative role.

We finally emphasize that the destructive interference due to the coherent
feedback control is unidirectional determined by the propagation direction of
the quantum field which is easier to be realized and controlled in contrast to
that due to a cavity directly coupled to an optomechanical system~\cite{84,85}%
. The cavity directly coupled scheme requires both individual addressability
of each cavity and large coupling strength between each other through spatial
proximity which is still a huge challenge~\cite{Arka majumdar}. However, in
our proposed scheme, such two cavities are spatially separated by each other
and directly coupling is replaced by coherent-feedback control which liberates
such two cavities in space and eliminates the slashing requirement about
challenging individual addressability and large coupling strength. Different
from the single driving way by applying the driving field to the linear cavity
in coupled-cavities scheme, our scheme can support another driving way, e.g.,
driving the controller cavity, which can also present strong photon blockade
effect even in the optomechanical single-photon weak-coupling regime. Another
significant difference is that both the controlled linear cavity and
optomechanical controller cavity can emerge photon blockade simultaneously in
contrast to the situation that only the linear cavity can emerge photon
blockade and optomechanical system can not in coupled-cavity scheme.

\section{CONCLUSION}

In summary, we have studied a system in which a linear cavity is coherently
controlled by an optomechanical system through a closed feedback loop. The
linear dynamics of the controlled cavity is nonlinearized by the
optomechanical controller via the so-called feedback nonlinearization. The
nonlinear controller using optomechanical systems and the coherent-feedback
loop are the core of this strategy. Such coherent-feedback strategy can both
liberates the controlled cavity and the optomechanical system in space and
eliminates the slashing requirement about individual addressability or large
coupling strength to achieve strong photon blockade in optomechanical
weak-coupling regime. Moreover, we find that the coherent feedback control
induced nonlinearity can be used to demonstrate some other interesting
nonlinear optical phenomena.

In the semiclassical regime, we found that optical bistability phenomenon in
the controller cavity induced by the optomechanical interaction can be
transferred to the controlled linear cavity through coherent feedback control.
In particular, we found that the threshold value of the bistability can be
significantly suppressed to zero through the coherent feedback control in
contrast to finite threshold value of the bistability in pure optomechanical
systems~\cite{80}.

In the quantum regime, we study the statistical properties of photons of the
controller cavity field and the controlled cavity field with two different
driving strategies. We found that photon blockade both in the controller
cavity and the controlled cavity is very similar under such two different
driving methods when the output damping rates of the controlled cavity is
comparable to those of the controller cavity. However, this similar effect can
be broken when the output damping rates of the controlled cavity are much
bigger than those of the controller cavity. Particularly, quite opposite
quantum nonlinear behaviors, e.g., tunneling and blockade, are discussed in
the region around the point $\Delta/\chi=2$. Photon blockade can be observed
when the coherent-coupling strengths $\sqrt{\kappa\gamma}$ and $\sqrt
{\kappa_{\mathrm{f}}\gamma}$ are enhanced no matter the driving is applied to
controlled cavity or the controller cavity. Moreover, photon blockade can
still happen even in the weak optomechanical coupling regime due to the
destructive quantum interference induced by coherent feedback control. It is
worth noting that the cavities's losses are actually crucial for achieving
photon blockade in such coherent-feedback approach in contrast to photon
blockade systems introduced so far in which the cavities' losses always play a
negative role.

We hope that our study can provide a controllable way to engineer strong
quantum nonlinearity, such as the control of photon transmission through a
linear cavity by using a optomechanical controller, achieving strong photon
blockade under weak optomechanical coupling condition and serving as
single-photon devices. We also hope that such proposed design can have
potential applications in quantum state engineering, quantum computing and
quantum communication.
\section{Acknowledgement}
Y.X.L. is supported by the National Natural Science Foundation
of China under Grant Nos. 61025022, 61328502, 91321208. J.Z. is supported by the National Natural Science Foundation
of China under Grant Nos. 61174084, 61134008. Y.X.L. and J.Z. are supported by the National Basic Research Program of China 973 Program under Grant No. 2014CB921401, the Tsinghua University Initiative Scientific Research Program, and the Tsinghua
National Laboratory for Information Science and Technology (TNList) Cross-discipline Foundation.


\begin{thebibliography}{99}                                                                                               %


\bibitem {1}T.~A. Fulton and G.~J. Dolan, Phys. Rev. Lett.~\textbf{59}, 109 (1987).

\bibitem {2}M.~A. Kastner, Reviews of Modern Physics~\textbf{64}, 849 (1992).

\bibitem {3}K.~K. Likharev, Proc. IEEE~\textbf{87}, 606 (1999).

\bibitem {4}A. Imamo\={g}lu, H. Schmidt, G. Woods, and M. Deutsch, Phys. Rev.
Lett.~\textbf{79}, 1467 (1997).

\bibitem {5}P. Grangier, D.~F. Walls, and K.~M. Gheri, Comment on
\textquotedblleft Strong interacting photons in a nonlinear
cavity\textquotedblright, Phys. Rev. Lett.~\textbf{81}, 2833 (1998).

\bibitem {6}F.-Y. Hong and S.-J. Xiong, Phys. Rev. A~\textbf{78}, 013812 (2008).

\bibitem {7}B. Dayan, A.~S. Parkins, T. Aoki, E.~P. Ostby, K.~J. Vahala, and
H.~J. Kimble, Science~\textbf{319}, 1062 (2008).

\bibitem {8}T. Aoki, A.~S. Parkins, D. J. Alton, C. A. Regal, B. Dayan, E.
Ostby, K.~J. Vahala, and H. J. Kimble, Phys. Rev. Lett.~\textbf{102}, 083601 (2009).

\bibitem {9}P. Michler, A. Kiraz, C. Becher, W.~V Schoenfeld, P.~M. Petroff,
L. Zhang, E. Hu, and A. Imamoglu, Science~\textbf{290}, 2282 (2000).

\bibitem {10}L. Zhou, L.~P. Yang, Y. Li, and C.~P. Sun, Phys. Rev.
Lett.~\textbf{111}, 103604 (2013).

\bibitem {11}S. Rosenblum, S. Parkins, and B. Dayan, Phys. Rev. A~\textbf{84},
033854 (2011).

\bibitem {12}A. Faraon, I. Fushman, D. Englund, N. Stoltz, P. Petroff, and J.
Vu\v{c}kovi\'{c}, Nat. Phys.~\textbf{4}, 859 (2008).

\bibitem {13}Q.~A. Turchette, C.~J. Hood, W. Lange, H. Mabuchi, and H.~J.
Kimble, Phys. Rev. Lett.~\textbf{75}, 25 (1995).

\bibitem {14}D. Schrader, I. Dotsenko, M. Khudaverdyan, Y. Miroshnychenko, A.
Rauschenbeutel, and D. Meschede, Phys. Rev. Lett.~\textbf{93}, 150501 (2004).

\bibitem {16}G.~J. Milburn, Phys. Rev. Lett.~\textbf{62}, 18 (1989).

\bibitem {17}H.~M. Gibbs, \textit{Optical Bistability: Controlling Light with
Light} (Academic, Orlando, 1985).

\bibitem {18}J.~L. O'Brien, Science~\textbf{318}, 1567 (2007).

\bibitem {19}K.~M. Birnbaum, A. Boca, R. Miller, A.~D. Boozer, T.~E. Northup,
and H.~J. Kimble, Nature~\textbf{436}, 87 (2005).

\bibitem {20}D. Englund, A. Majumdar, A. Faraon, M. Toishi, N. Stoltz, P.
Petroff, and J. Vu\v{c}kovi\'{c}, Phys. Rev. Lett.~\textbf{104}, 073904 (2010).

\bibitem {21}C. Lang, D. Bozyigit, C. Eichler, L. Steffen, J.~M. Fink, A.~A.
Abdumalikov, M. Baur, S. Filipp, M.~P. da Silva, A. Blais, and A. Wallraff,
Phys. Rev. Lett.~\textbf{106}, 243601 (2011).

\bibitem {22}A.~J. Hoffman, S.~J. Srinivasan, S. Schmidt, L. Spietz, J.
Aumentado, H.~E. T\"{u}eci, and A.~A. Houck, Phys. Rev. Lett.~\textbf{107},
053602 (2011).

\bibitem {23}T.~J. Kippenberg and K.~J. Vahala, Science~\textbf{321}, 1172 (2008).

\bibitem {24}F. Marquardt and S. M. Girvin, Physics~\textbf{2}, 40 (2009).

\bibitem {Clerk}A.~A. Clerk, M.~H. Devoret, S.~M. Girvin, F. Marquardt, and
R.~J. Schoelkopf, Rev. Mod. Phys.~\textbf{82}, 1155 (2010).

\bibitem {25}M. Aspelmeyer, P. Meystre, and K. Schwab, Phys. Today~\textbf{65}%
, 29 (2012).

\bibitem {26}M. Aspelmeyer, T.~J. Kippenberg, and F. Marquardt, arXiv:1303.0733.

\bibitem {27}V.~B. Braginsky, S.~E. Strigin, and S.~P. Vyatchanin, Phys. Lett.
A~\textbf{305}, 111 (2002).

\bibitem {28}J.~M. Courty, A. Heidmann, and M. Pinard, Phys. Rev. Lett.
\textbf{90}, 083601 (2003).

\bibitem {29}T. Corbitt, Y. Chen, E. Innerhofer, H. M\"{u}ler-Ebhardt, D.
Ottaway, H. Rehbein, D. Sigg, S. Whitcomb, C. Wipf, and N. Mavalvala, Phys.
Rev. Lett. \textbf{98}, 150802 (2007).

\bibitem {30}W. Marshall, C. Simon, R. Penrose, and D. Bouwmeester, Phys. Rev.
Lett. \textbf{91}, 130401 (2003).

\bibitem {31}D. Rugar, R. Budakian, H.~J. Mamin, and B.~W. Chui,
Nature~\textbf{329} (2004).

\bibitem {32}P. Rabl, S.~J. Kolkowitz, F. H. L. Koppens, J. G. E. Harris, P.
Zoller, and M.~D. Lukin, Nat. Phys.~\textbf{6}, 602 (2010).

\bibitem {33}S. Gigan, H.~R. B\"{o}hm, M. Paternostro, F. Blaser, G. Langer,
J.~B. Hertzberg, K.~C. Schwab, D. B\"{a}uerle, M. Aspelmeyer, and A.
Zeilinger, Nature~\textbf{444}, 67 (2006).

\bibitem {34}O. Arcizet, P.-F. Cohadon, T. Briant, M. Pinard, and A. Heidmann,
Nature~\textbf{444}, 71 (2006).

\bibitem {35}A. Schliesser, P. Del'Haye, N. Nooshi, K. Vahala, and T.
Kippenberg, Phys. Rev. Lett.~\textbf{97}, 243905 (2006).

\bibitem {36}I. Wilson-Rae, N. Nooshi, W. Zwerger, and T. J. Kippenberg, Phys.
Rev. Lett.~\textbf{99}, 093901 (2007).

\bibitem {37}F. Marquardt, J. Chen, A. Clerk, and S. Girvin, Phys. Rev. Lett.
\textbf{99}, 093902 (2007).

\bibitem {38}J. Teufel, J. Harlow, C. Regal, and K. Lehnert, Phys. Rev. Lett.
\textbf{101}, 197203 (2008).

\bibitem {39}J.~D. Thompson, B.~M. Zwickl, A.~M. Jayich, F. Marquardt, S.~M.
Girvin, and J.G.E. Harris, Nature~\textbf{452}, 72 (2008).

\bibitem {40}A. Schliesser, R. Rivi\`{e}re, G. Anetsberger, O. Arcizet, and
T.~J. Kippenberg, Nat. Phys.~\textbf{4}, 415 (2008).

\bibitem {41}S. Gr\"{o}blacher, J.~B. Hertzberg, M.~R. Vanner, G.~D. Cole, S.
Gigan, K.~C. Schwab, and M. Aspelmeyer, Nat. Phys.~\textbf{5}, 485 (2009).

\bibitem {42}T. Rocheleau, T. Ndukum, C. Macklin, J.~B. Hertzberg, A.~A.
Clerk, and K.~C. Schwab, Nature~\textbf{463}, 72 (2010).

\bibitem {43}Y.~S. Park and H. Wang, Nat. Phys.~\textbf{5}, 489 (2009).

\bibitem {44}A. Schliesser, O. Arcizet, R. Rivi\`{e}re, G. Anetsberger, and
T.~J. Kippenberg, Nat. Phys.~\textbf{5}, 509 (2009).

\bibitem {45}J.~D. Teufel, T. Donner, D. Li, J.~W. Harlow, M.~S. Allman, K.
Cicak, A.~J. Sirois, J.~D. Whittaker, K.~W. Lehnert, and R.~W. Simmonds,
Nature ~\textbf{475}, 359 (2011).

\bibitem {46}J. Chan, T.P.M. Alegre, A.~H. Safavi-Naeini, J.~T. Hill, A.
Krause, S. Gr\"{o}blacher, M. Aspelmeyer, and O. Painter, Nature~\textbf{478},
89 (2011).

\bibitem {47}E. Verhagen, S. Del\'{e}glise, S. Weis, A. Schliesser, and T.~J.
Kippenberg, Nature~\textbf{482}, 63 (2012).

\bibitem {48}T. Carmon, H. Rokhsari, L. Yang, T. J. Kippenberg, and K. J.
Vahala, Phys. Rev. Lett.~\textbf{94}, 223902 (2005).

\bibitem {49}T. J. Kippenberg, H. Rokhsari, T. Carmon, A. Scherer, and K. J.
Vahala, Phys. Rev. Lett.~\textbf{95}, 033901 (2005).

\bibitem {50}J. D. Teufel, D. Li, M. S. Allman, K. Cicak, A.~J. Sirois, J.~D.
Whittaker, and R.~W. Simmonds, Nature~\textbf{471}, 204 (2011).

\bibitem {51}G. S. Agarwal and S. Huang, Phys. Rev. A~\textbf{81}, 041803(R) (2010).

\bibitem {52}S. Huang and G. S. Agarwal, Phys. Rev. A~\textbf{83}, 043826 (2011).

\bibitem {53}S. Weis, R. Rivi\`{e}re, S. Del\'{e}glise, E. Gavartin, O.
Arcizet, A. Schliesser, and T. J. Kippenberg, Science~\textbf{330}, 1520 (2010).

\bibitem {54}A.~H. Safavi-Naeini, T.~P. Mayer Alegre, J. Chan, M. Eichenfield,
M. Winger, Q. Lin, J.~T. Hill, D. E. Chang, and O. Painter,
Nature~\textbf{472}, 69 (2011).

\bibitem {55}D. Vitali, S. Gigan, A. Ferreira, H. R. B\"{o}hm, P. Tombesi, A.
Guerreiro, V. Vedral, A. Zeilinger, and M. Aspelmeyer, Phys. Rev.
Lett.~\textbf{98}, 030405 (2007).

\bibitem {56}M. J. Hartmann and M. B. Plenio, Phys. Rev. Lett.~\textbf{101},
200503 (2008).

\bibitem {57}K. Stannigel, P. Rabl, A.~S. S{\o }rensen, P. Zoller, and M.~D.
Lukin, Phys. Rev. Lett.~\textbf{105}, 220501 (2010).

\bibitem {58}S. Gr\"{o}lacher, K. Hammerer, M.~R. Vanner, and M. Aspelmeyer,
Nature~\textbf{460}, 724 (2009).

\bibitem {59}L. Tian and H. Wang, Phys. Rev. A \textbf{82}, 053806 (2010); L.
Tian, Phys. Rev. Lett.~\textbf{108}, 153604 (2012).

\bibitem {60}Y.~D. Wang and A.~A. Clerk, Phys. Rev. Lett.~\textbf{108}, 153603 (2012).

\bibitem {61}S.~A. McGee, D. Meiser, C.~A. Regal, K.~W. Lehnert, and M.~J.
Holland, Phys. Rev. A~\textbf{87}, 053818 (2013).

\bibitem {62}J.~T. Hill, A.~H. Safavi-Naeini, J. Chan, and O. Painter, Nat.
Commun.~\textbf{3}, 1196 (2012).

\bibitem {HLWang}C. Dong, V. Fiore, M.~C. Kuzyk, and H. Wang,
Science~\textbf{338}, 1609 (2012)

\bibitem {63}J. Bochmann, A. Vainsencher, D.~D. Awschalom, and A.~N. Cleland,
Nat. Phys.~\textbf{9}, 712 (2013).

\bibitem {Lehnert}R.~W. Andrews, R.~W. Peterson, T.~P. Purdy, K. Cicak, R.~W.
Simmonds, C.~A. Regal, and K.~W. Lehnert, Nature Phys.~\textbf{10}, 321 (2014).

\bibitem {Gong}Z.~R. Gong, H. Ian, Y.~X. Liu, C.~P. Sun, and F. Nori, Phys.
Rev. A~\textbf{80}, 065801 (2009).

\bibitem {80}S. Aldana, C. Bruder, and A. Nunnenkamp, Phys. Rev.
A~\textbf{88}, 043826 (2013).

\bibitem {Lu}X.~Y. L\"{u}, W.~M. Zhang, S. Ashhab, Y. Wu, and F. Nori,
Sientific Rep.~\textbf{3}, 2943 (2013).

\bibitem {AgarwalPRA85}G.~S. Agarwal and S. Huang, Phys. Rev. A~\textbf{85},
021801(R) (2012).

\bibitem {64}P. Rabl, Phys. Rev. Lett.~\textbf{107}, 063601 (2011).

\bibitem {65}A. Nunnenkamp, K. B{\o }rkje, and S.~M. Girvin, Phys. Rev.
Lett.~\textbf{107}, 063602 (2011).

\bibitem {83}X.~W. Xu, Y.~J. Li, and Y.~X. Liu, Phys. Rev. A~\textbf{87},
025803 (2013).

\bibitem {Jieqiao1}J.~Q. Liao, H.~K. Cheung, and C.~K. Law, Phys. Rev.
A~\textbf{85}, 025803 (2012).

\bibitem {Jieqiao2}J.~Q. Liao and F. Nori, Phys. Rev. A~\textbf{88}, 023853 (2013).

\bibitem {66}K. Stannigel, P. Komar, S. J. M. Habraken, S.~D. Bennett, M.~D.
Lukin, P. Zoller, and P. Rabl, Phys. Rev. Lett.~\textbf{109}, 013603 (2012).

\bibitem {67}M. Ludwig, A.~H. Safavi-Naeini, O. Painter, and F. Marquardt,
Phys. Rev. Lett.~\textbf{109}, 063601 (2012).

\bibitem {68}X.~W. Xu and Y.~J. Li, J. Opt. B: At. Mol. Opt. Phys.~\textbf{46}%
, 035502 (2013).

\bibitem {Arka majumdar}A. Majumdar, M. Bajcsy, A. Rundquist and J.
Vu\v{c}koci\'{c}, Phys. Rev. Lett.~\textbf{108}, 183601 (2012).

\bibitem {78}J. Zhang, R.~B. Wu, Y.~X. Liu, and C.~W. Li, IEEE Trans. Automat.
Contr. ~\textbf{57}, 1997 (2012).

\bibitem {Z.P.L}Z.~P. Liu, H. Wang, J. Zhang, Y.~X. Liu, R.~B. Wu, C.~W. Li,
and F. Nori, Phys. Rev. A~\textbf{88}, 063851 (2013).

\bibitem {Feedback}H. Wiseman and G. Milburn, Physical Review A \textbf{49},
4110 (1994).

\bibitem {69}S. Lloyd, Phys. Rev. A~\textbf{62}, 022108 (2000); R.~J. Nelson,
Y. Weinstein, D. Cory, and S. Lloyd, Phys. Rev. Lett.~\textbf{85}, 3045 (2000).

\bibitem {70}H. Mabuchi, Phys. Rev. A~\textbf{78}, 032323 (2008); M. Armen, J.
Au, J. Stockton, A. Doherty, and H. Mabuchi, Phys. Rev. Lett. \textbf{89},
133602 (2002).

\bibitem {71}G.~G. Gillett, R.~B. Dalton, B.~P. Lanyon, M.~P. Almeida, M.
Barbieri, G.~J. Pryde, J.~L. O'Brien, K.~J. Resch, S.~D. Bartlett, and A.~G.
White, Phys. Rev. Lett. \textbf{104}, 080503 (2010).

\bibitem {72}M.~R. James, H.~I. Nurdin, and I.~R. Petersen, IEEE Trans.
Automat. Contr.~\textbf{53}, 1787 (2008).

\bibitem {73}J. Gough and M.~R. James, IEEE Trans. Automat. Contr.~\textbf{54}%
, 2530 (2009).

\bibitem {74}M. Yanagisawa and H. Kimura, IEEE Trans. Automat.
Contr.~\textbf{48}, 2107 (2003).

\bibitem {75}Z. Zhou, C. Liu, Y. Fang, J. Zhou, R.~T. Glasser, L. Chen, J.
Jing, and W. Zhang, Appl. Phys. Lett.~\textbf{101}, 191113 (2012).

\bibitem {76}J. Kerckhoff and K. W. Lehnert, Phys. Rev. Lett.~\textbf{109},
153602 (2012).

\bibitem {FeedbackExp1}A Kubanek, M. Koch, C. Sames, A Ourjoumtsev, P. W. H.
Pinkse, K. Murr, and G. Rempe, Nature \textbf{462}, 898 (2009).

\bibitem {FeedbackExp2}H. Yonezawa, D. Nakane, T.~A. Wheatley, K. Iwasawa, S.
Takeda, H. Arao, K. Ohki, K. Tsumura, D.~W. Berry, T.~C. Ralph, H.~M. Wiseman,
E.~H. Huntington, and A. Furusawa, Science~\textbf{337}, 1514 (2012).

\bibitem {FeedbackExp3}O. Arcizet, P.-F. Cohadon, T. Briant, M. Pinard, and A.
Heidmann, Nature \textbf{444}, 71 (2006); D. Kleckner and D. Bouwmeester,
Nature~\textbf{444}, 75 (2006).

\bibitem {77}J. Zhang, Y.~X. Liu, R.~B. Wu, K. Jacobs, and F. Nori, Phys. Rev.
A~\textbf{87}, 032117 (2013).

\bibitem {81}D.~F. Walls and G.~J. Milburn, \textit{Quantum Optics}
(Spribger-Verlag, Heidelberg, 2008).

\bibitem {82}R.~R. Puri, Mathematical Methods of Quantum Optics
(Spribger-Verlag, Berlin, 2001).

\bibitem {84}M. Bamba, A. Imamo\u{g}lu, I. Carusotto, and C. Ciuti, Phys. Rev.
A~\textbf{83}, 021802 (2011).

\bibitem {85}T.C.H. Liew and V. Savona, Phys. Rev. Lett.~\textbf{104}, 183601 (2010).
\end{thebibliography}
\end{document}